# A Structure-Centric View of Protein Evolution, Design and Adaptation

Eric J. Deeds* and Eugene I. Shakhnovich†

*Department of Systems Biology, Harvard Medical School, 200 Longwood Avenue, Boston, MA 02138, USA*

*†Department of Chemistry and Chemical Biology, Harvard University, 12 Oxford Street, Cambridge, MA 02138, USA*

# Abstract

Proteins, by virtue of their central role in most biological processes, represent one of the key subjects of the study of molecular evolution. Inherent to the indispensability of proteins for living cells is the fact that a given protein can adopt a specific three-dimensional shape that is specified solely by the protein's sequence of amino acids. Over the past several decades, structural biologists have demonstrated that the array of structures that proteins may adopt is quite astounding, and this has lead to a strong interest in understanding how protein structures change and evolve over time. In this review we consider a large body of recent work that attempts to illuminate this structure-centric picture of protein evolution. Much of this work has focused on the question of how completely new protein structures (i.e. new folds or topologies) are discovered by protein sequences as they evolve. Pursuant to this question of structural innovation has been a desire to describe and understand the observation that certain types of protein structures are far more abundant than others and how this uneven distribution of proteins implicates on the process through which new shapes are discovered. We consider a number of theoretical models that have been successful at explaining this heterogeneity in protein populations and discuss the increasing amount of evidence that indicates that the process of structural evolution involves the divergence of protein sequences and structures from one another. We also consider the topic of protein designability, which concerns itself with understanding how a protein's structure influences the number of sequences that can successfully fold into that structure. Understanding and quantifying the relationship between the physical feature of a structure and its designability has been a long-standing goal of the study of protein structure and evolution, and we discuss a number of recent advances that have yielded a promising answer to this question. Finally we review the relatively new field of protein structural phylogeny, an area of study in which information about the distribution of protein structures among different organisms is used to reconstruct the evolutionary relationships between them. Taken together, the work that we review presents an increasingly coherent picture of how these unique polymers have evolved over the course of life on Earth.

# Introduction

Understanding the evolution of proteins represents an important task in the study of molecular evolution. Proteins play a central role in almost every process in the cell, and it is clear that life as we know it could not exist without this functionally diverse class of macromolecules. The great functional capacity and importance of proteins largely stems from the remarkable ability of these polymers to adopt distinct 3-dimensional structures. As Anfinsen's experiment first demonstrated (1), protein structures are for the most part specified by their amino acid sequences, and the variety of protein structures that may be achieved by varying these sequences in a 20-letter alphabet is truly astounding (2-6). The strong correspondence between protein structure and function indicates that any attempt to understand how life has evolved to its present-day form must explain the changes in protein sequence and structure that have occurred over the past ~3.5 billion years.

The structure-centric view of protein evolution seeks, at its core, to understand the process whereby novel protein structures and folds are discovered (6). Although such discovery events are likely to be comparatively rare, they are of obvious and extreme importance to living organisms. New sequence-structure pairs can substantially vary enzymatic, regulatory and mechanical functions that may be achieved by proteins; entirely new folds and topologies can provide completely new chemistry and biology to a cell. The observation of structural diversity in the natural world is clear evidence that such structural innovation events have occurred over the course of evolution, and much work has been done to understand how this process occurs. As databases of sequence and structure information have grown in size it has become possible to understand structural diversity in a systematic way, posit models describing the progress of structural evolution, and test those models against quantitative features of the existing protein universe. Our analysis of this work begins with an overview of the various approaches to encoding and quantifying structural diversity and structural relationships, from earlier human-annotated databases of structural taxonomy to the more recent use of quantitative structural comparisons to build graph theoretic descriptions of the protein universe. These various representations are informed by, and in turn inform, the long-standing debate between two diametrically opposed paradigms for protein evolution: convergence and divergence. We review the current status of this debate and also look in detail at the performance of a wide variety of divergent models that have been put forth within the last 10 years to explain structural innovation.

  Although the mechanism by which sequences discover new types of protein structure is inherently interesting and important, the structure-centric view also looks to elucidate the design principles that underlie the correspondence between sequence, structure and function. To what extent do protein structures constrain the sequences that can fold into them? How does the level of constraint and the number of possible sequences vary from structure to structure? How functionally flexible are different structures? What signatures of these varying structural influences can be found in either structures or sequences? As the amount of available data has grown, answers to these fundamental questions are beginning to appear. In this chapter we review the long history of the "designability" principle, which is the idea that some structures inherently correspond to more sequences than others. Although this concept was first expressed in terms of the convergent paradigm of structural evolution, we explore the current view that designability may actually have a large impact on the nature of evolution even within a divergent paradigm. We also consider the various methods that have been proposed to quantify how designable a structure should be based solely on its physical features. We look at the implications that designability may have for the sizes of existing protein families, their thermodynamic stability, and their ability to perform diverse chemical and biological functions.

Finally, the structure-centric view seeks to understand the natural history of protein domains and build a comprehensive picture of both structural and organismal evolution. Questions of phylogeny have long been the domain of sequence analysis, but recent work has demonstrated that protein structures have a lot to tell us about the evolution of living systems beyond simple questions of their own evolution. Although this particular field is still very much in its infancy, we review the recent efforts that have attempted to parlay the burgeoning availability of structural information into improvements in phylogenetic understanding.

## Understanding the Protein Structural Universe

### Structural Classification

As mentioned above, the set of protein structures that populate modern databases exhibits an amazing level of diversity (2-7). It is in many ways instructive to think of this diversity as the protein analogue of the diversity of forms (or "structures") we observe for organisms in the natural world. In both cases the array of structures is at once intriguing and bewildering (the CATH wheel displayed in Figure 1, first introduced by Thornton and co-workers (5), captures this structural diversity quite nicely). Despite this diversity, careful analysis of organismal structures leads to the realization that there seem to be organizing principles and systematic relationships that govern the progression of shapes.

# Figure 1

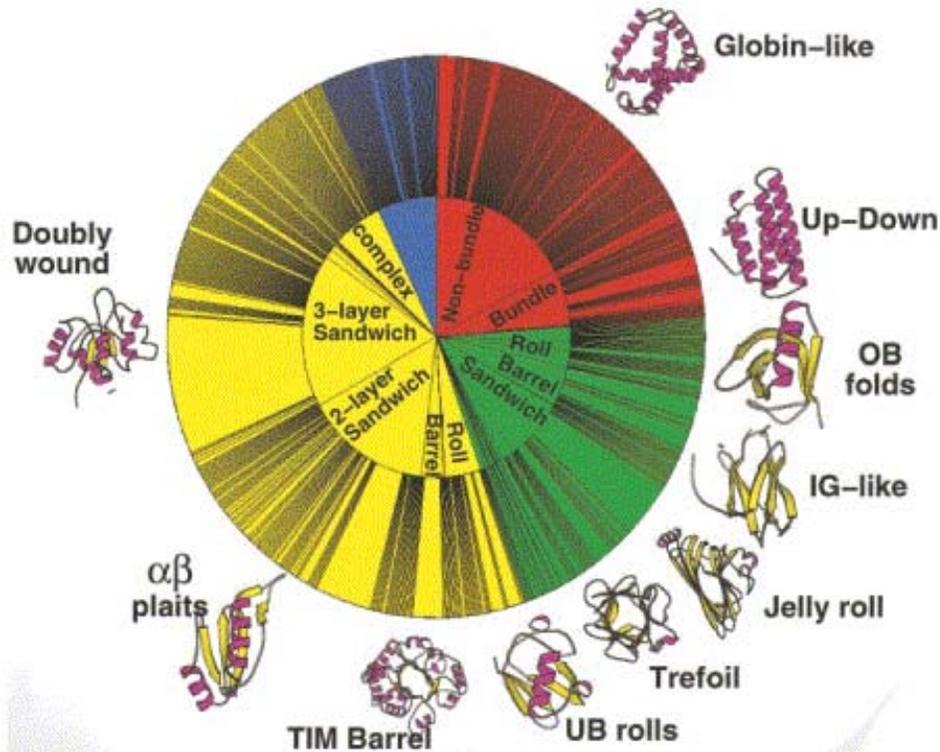

**Figure 1.** The CATH wheel, taken from (4). The diversity of 3-D shapes observed in the set of solved protein structures is exemplified by the structures surrounding the wheel; secondary structural content runs from all-$\alpha$ to $\alpha/\beta$ and, even within classes of secondary structural content, a wide variety of structures are observed. The sizes of the pie slices in the wheel essentially reflect the number of domains that display a particular architecture (i.e. TIM Barrel or IG-like).

For instance, it is clear that the different species of beetles, for all of their diversity, are more closely related to one another than they are to other insects; that is, organisms have certain structural characteristics that seem to define a set of hierarchical categories to which they naturally belong. In analogy to this taxonomy of organisms, early analysis of protein structural databases revealed that there were systematic variations in protein structure in which concepts of "homology" in sequence and structure make sense (3, 5, 6, 8). This work led to the critical idea that proteins could be categorized with respect to one another on the basis of their degree of similarity in sequence- and structure-based characteristics. This allows for the existing protein structural universe to be analyzed in systematic ways and is crucial for understanding how protein sequences and structures evolve (6).

Databases of structural classification are familiar to most protein scientists. Perhaps the most well known are SCOP (Structural Classification of Proteins) developed by Murzin and colleagues(3), CATH (Class Architecture Topology Homology) by Thornton and colleagues (7) and FSSP (Families of Structurally Similar Proteins) by Holm, Sander and colleagues (9). The two former databases (SCOP and CATH) are organized around similar taxonomic principles and involve some measure of human curation (much as taxonomy often relies on human judgments of similarity). FSSP is a fully computational and automated form of structural classification aimed at similar taxonomic goals (9, 10).

The basic unit of structural taxonomy (i.e. the "species" that are classified with respect to one another) is the protein domain. A domain is defined as an independently folding structural unit (6), and every protein structure consists of at least one structural domain. The fact that such domains are considered stable and foldable on their own indicates that they may represent convenient, independent evolutionary units and provides an appropriate set of species from which the taxonomy may be built. Although the definition of these domains is a non-trivial problem (3, 6, 8), these structures are well accepted as the most appropriate and basic unit in protein classification.

Domains from different proteins are grouped together in the first (several) levels of classification on the basis of sequence similarity. Although all protein classification systems are inherently structural in nature, the fact that sequence homologs are also close structural homologs informs these levels of classification (3, 5-7, 11). Sequence homologs are certainly very closely related in evolutionary terms, and it is clear that orthologs and paralogs of a given domain share a common sequence ancestor. These sets of similar sequences are grouped together as a domain family in the hierarchical classification (3, 6, 9). The next level of classification is that of a "fold," which represents a cluster of domain families that exhibit similar structures. At the fold level there is ostensibly little or no sequence similarity between members of the different sequence families that make up the fold, that is, different members of a fold exhibit structural similarity in the absence of sequence similarity.

Given the large amount of work that has been done to understand the statistical and evolutionary meaning of sequence overlap (3, 5, 6, 12, 13), classification of sequences into domain families is fairly straightforward. Classification of those structures into folds is not so simple. In the case of both SCOP and CATH, these classifications have at least some human component; for instance the SCOP database structural classifications rely heavily on the structural intuition of Murzin himself (3, 6, 7). Such intuitive structural classifications are not necessarily problematic; indeed, they represent a very appealing scheme since the very intuition for structural similarity that suggested the existence and evolutionary utility of a structural taxonomy is employed to define it. Human classification does, however, introduce problems if one wishes to reliably quantify the structural relationships between domains. It is clear that all pairwise relationships between (family) domain structures at the fold level are not equivalent; some TIM barrel families look more similar to each other than other TIM barrel families. To overcome this difficulty, Holm and Sander introduced a completely automated method for defining structural distances (10), defining protein domains (8) and classifying them with respect to one another (9). The structural comparison algorithm they employ, DALI, is based on measuring the similarity between the inter-residue distance matrices (i.e. contact maps) that characterize two different structures (10). Structural similarity is quantified as the DALI Z-score, which represents the statistical significance of a structural overlap. The FSSP database classifies structures into a level equivalent to the "fold" using DALI in a completely automatic way (9). Given the fact that DALI produces structural similarity scores commensurate with our expectations for structural similarity, it is not surprising to find that the correspondence between SCOP, CATH and FSSP classification assignments is fairly high (14, 15). Although higher levels of structural taxonomy exist (i.e. structural class such as $\alpha/\beta$), for the purposes of understanding the nature of the protein structural universe the fold component of the hierarchy has provided perhaps the most insight.

**Distribution of Folds and the Convergent/Divergent Debate**

The classification schemes discussed above, regardless of their reliance on human or computational structural homology measurements, provide very similar results in terms of the distribution of protein structures within protein structural space. Some protein folds contain many, many more members than other folds (see, for instance, the differences in the sizes of the "pie slices" in the CATH wheel in Figure 1)(5, 6). For instance the number of domains that are classified in the TIM barrel fold is much larger than the number classified in the Phosphoglycerate Kinase fold (3). Explaining the source of this heterogeneity has been the subject of much study and has lead to the proposal of two opposing views of protein structural evolution: convergence and divergence.

Perhaps the earliest proposal aimed at explaining the heterogeneity in fold and family sizes was the idea of designability. This concept, attributed to Finklestein and colleagues (16-18), proposes that some protein structures can be coded for by many distinct sequences and are thus more designable than those structures that can be coded for by comparatively few sequences. Although the idea of designability is in no way incompatible with either convergent or divergent pictures of protein evolution, the convergent argument relies heavily on it to describe the empirical differences in fold sizes (6, 16, 17). The argument runs like this: suppose that the total number of possible polypeptide structures (i.e. the set of all possible folds) is fairly small. Two domain sequences evolving independently of one another discover these different folds at random, and the proportion of times a given fold is discovered is related quite naturally to the designability of that fold. Under this hypothesis the observation that two domain families share significant topological and structural features (and as such are classified together in one fold) indicates that the two sequences converged to the same moderately or highly designable structure.

As more and more structural information became available and it became possible to start quantifying the amount of heterogeneity in fold sizes, an alternative divergent picture of protein evolution emerged (6). This picture holds that the number of protein sequence families and protein structural folds is large and the structures we observe today are the result of evolutionary dynamics and not some inherent feature of the fold. This paradigm of structural evolution posits that members of a given fold share a common ancestor that exhibited the topological features that define the fold. In this paradigm, pairwise structural similarity in the absence of sequence similarity is in fact an indicator of ancient sequence homology, and all members of a fold derive from a single common ancestor (6). Both the divergent and convergent paradigms have their adherents, but in recent years the divergent picture has gained considerably in popularity(6). At this point we will consider the arguments that characterize both sides and then proceed to review the salient features of the protein structural universe that have been used to support each claim.

Until recently the most common arguments against the convergent picture revolved around the fact that the assumptions underlying it simply do not make sense(6). The size of sequence space is massive (on the order of $10^{130}$ or more sequences depending on the sequence length), and so it is clear that the extent of sequence exploration represented by currently existing domains is a very tiny fraction of all sequence possibilities. From this argument it is easy to see that, even under extreme selective pressures, sequences whose common ancestor is in the very distant past (i.e. sequences that are unrelated to one another) do not have a very high likelihood of convergently discovering sequences with appreciable similarity. Even though the concept of "equilibrium" in sequence space is thus a ludicrous one, the convergent picture of structural evolution does not rely on small sequence spaces. It does, however, rely on the assumption that the number of structural possibilities is small and that the probability of finding a particular fold in sequence space (which is a function of the designability of that fold) is distributed roughly equivalently in all regions of sequence space. Think of a random walk on an infinitely large chessboard; regardless of the fact that you cannot walk the entire board in finite time, the proportion of times you step on a white or black square depends only on the fraction of squares that are white and black. Similarly, if the total number of folds is small and regularly distributed it is not difficult to imagine independent, convergent discovery of two domains in the same fold by two very distantly related sequences.

The total number of possible folds is much more difficult to estimate than the total number of sequences. Survey of the existing protein universe reveals a total of ~800 folds (3), but as this represents only the existing sample of solved structures it is difficult to extrapolate from this number to the actual size of the entire fold universe. Attempts aimed at estimating the number of folds that have evolved on earth vary from 1000 to 10,000 total folds, a discrepancy that highlights the difficulties inherent in these kinds of estimations (6, 19-24). The protein folding and structure prediction problem remains unsolved for arbitrary sequences (25), and until the problem is solved it is unlikely that much reliable progress will be made on estimating the size of either the overall structural universe or the evolved subset at the fold level. The overall size is likely to be a very large number (6), but there is currently no way to reliably *demonstrate* that it is a large enough number to render the convergent scenario unlikely.

The increase in the amount of structural information contained in public databases over the past 10 years has allowed for more systematic and quantitative analysis of the heterogeneity in fold sizes. Given a fairly large number of solved protein structures it is possible to calculate the probability of observing a fold with a certain number of constituent families. The fold size distribution in SCOP and other databases has been calculated at different times by a number of different groups, and in each case it has been shown that this distribution is well-described by a scaling law, i.e. if $m$ is the number of families in a fold then $p(m) \sim m^{-\mu}$ (6, 26-28) (one example of this observation, taken from the work of Gerstein and coworkers (26), is shown in Figure 2A and 2B). These observations clearly demonstrate the fold size heterogeneity mentioned in passing above: in such distributions folds of size one (known as "orphans") dominate the population but folds with 100s of member domains may also be found (6, 26-28). This "scale-free" fold size distribution provided perhaps the first global, quantitative feature of the evolved protein structural universe against which models of protein structural evolution could be tested.

A number of divergent models, collectively named the Birth, Death and Innovation (BDIM) models by Koonin and coworkers (6, 26, 29), were proposed to explain this behavior (for a schematic of one such model taken from (6), see Figure 2C). These models are based on the principle that specific sequences within domain families are sometimes duplicated and, through a sufficient number of mutations, found new domain families of their own. A newly "born" domain is always of the same fold as the parent domain, and so duplication of existing domains increases that fold's representation and consequently its size. Domain loss (or "death") mechanisms remove sequence families from existence and reduce a fold's size. Innovation is treated as a source that seeds new folds, and in a divergent model they basically represent the constant generation of brand new topologies from the set of all duplication events. In these models the birth

**Figure 2**

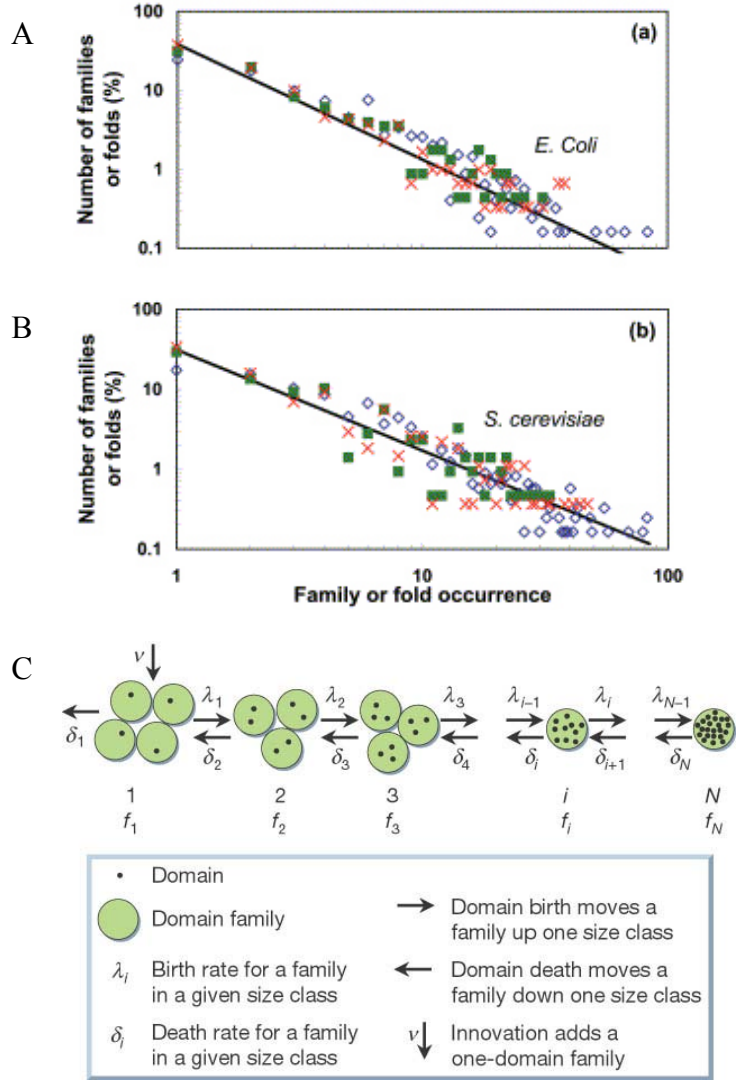

**Figure 2.** The Fold-Size distribution in organisms: observations and models. (A) Taken from (27), the distribution of SCOP fold sizes in *Escherichia coli* is indicated by the Xs. The x-axis represents the size of a given fold and the y-axis represents the probability of find a fold of that size in the genome of this particular organism. The other symbols in the graph represent other types of data, such as familizy sizes. The straight line indicates a power-law fit. (B) A plot similar to that in A (again taken from (27)) but for Saccharomyces cerevisiae. (C) A schematic of the BDIM model as a representative for divergent models that can reproduce the distributions in B and C. Taken from (6),

and death rates depend on the size of the fold (as one would expect) and can be varied independently of one another. In a certain parameter regime one finds that the distribution of fold sizes is indeed scale free. The trend in the parameters that give rise to this behavior are, in general, quite reasonable: they predict that the death rate roughly equals the birth rate for large folds but is much larger than the birth rate for small ones (6, 28). This could easily arise from the fact that larger folds are older than younger folds in BDIM models; although it is not explicitly modeled, the families in older folds will have had more time to accrue orthologs and paralogs and thus have a greater number of sequence representatives, making complete loss of that family from the evolved protein universe less likely.

The understanding of protein structural evolution that underlies the BDIM models, i.e. that of fold and family expansion under structural constraints, was explored in greater theoretical detail by Dokholyan and Shakhnovich (30). In this work they posited that protein evolution could be understood in terms of a "free energy landscape" in both sequence and structure space. Local exploration of sequence-structure pockets (which correspond to local minima on the evolutionary landscape) occurs on some timescale and represents the diffusion of orthologs and paralogs with respect to one another within this pocket. The pocket itself is defined by a key set of residues that are constrained to certain amino acids in order for that set of sequences to support folding into a given structure, a fact that results in the conservation of specific amino acids or amino acid types at certain positions within the sequence family (30). On a separate time scale, some sequences cross "barriers" in this landscape and seed new local minima. These local minima are unrelated from the standpoint of sequence comparison. The new sequence pocket may be subsequently explored on a shorter timescale with certain residues constrained. Sometimes these transitions result in structures that are similar to the original structure (i.e. duplication events that serve to increase fold sizes in the BDIM models). In this case, comparison of the two sequence pockets demonstrates that the *identity* of the conserved residues differs between the two but the structural similarity is maintained because the relative *positions* of these conserved residues do not change. In other cases structural similarity is not maintained and a brand new fold is discovered (corresponding to the "innovation" step in BDIM-type models) (30). Dokholyan and Shakhnovich explored a model of protein evolution involving real protein structures and found that those residues with low substitution rates in their model tended to have low "Conservatism of Conservatism" (CoC) entropies (30, 31). The CoC quantity considers families of sequences that belong to the same fold and identifies positions in those sequence families that are highly conserved within families (i.e. have low sequence variance) and tend to be highly conserved in the set of families in the fold (i.e. positions that have low sequence entropy in many families within the fold) (30, 31).

Although these results are encouraging from the standpoint of divergent structural evolution, the existence of a divergent model that explains some feature of the protein structural universe cannot be taken as evidence against the convergent hypothesis. In this case, the convergent hypothesis has only to assert that the underlying distribution of fold designabilities follows a scaling law itself. This assumption is difficult to prove, but early analysis of the designabilities of lattice polymer structures (in this case all compact conformations of 27-mers on the cubic lattice) (32) demonstrated that a power-law distribution of designabilities existed in this model system and thus was not out of the question for protein structures. Also, within a convergent picture, designability constraints introduced by structural characteristics of a fold could easily impose conservation patterns similar to those observed in divergent simulations in the absence of divergent mechanisms.

**The Protein Domain Universe Graph**

The next insight into the protein structural evolution on this scale involved looking at protein structures at greater (although still quite coarse-grained) resolution. As mentioned above it is clear that binning all structures of the "TIM barrel" topology together as one group, however natural, tends to ignore the existence of structural differences within a fold despite the fact that some differences are fairly clear (3). Moving to that level of description requires not only a method of measuring those pairwise relationships (which is impossible in databases such as SCOP and CATH due to their inherently binned and taxonomic organization) and a method of representing them in order to understand their implications for the protein structural universe. The former was provided by FSSP in the form of the DALI Z-score; the latter was provided by a graph theoretic representation of protein structural similarity. Dokholyan and coworkers combined these concepts, resulting in a representation of existing protein structures that is called the Protein Domain Universe Graph (PDUG) (33).

Although the protein structural universe had been discussed in terms of network theory (6, 29), the PDUG represented the first truly systematic application of this methodology to distributions of protein structural similarity. In the PDUG, nodes are taken to be domain families, and the structural comparisons that form the basis of the edges on the graph are computed using single structural representatives from these families. Given the strong correspondence in structure between related sequences, the choice of representative from each family should not matter in terms of the nature of the graph. The DALI Z-score of structural overlap between every pair of these representatives is employed to define the edges on the graph. To transform the weighted graph (in which every edge is, in a sense, "labeled" with the Z-score that defines it) into an unweighted graph in which every edge is treated equivalently, one defines a cutoff parameter $Z_{min}$ such that unweighted edges are only placed between domains $i$ and $j$ if $Z_{ij} \geq Z_{min}$ (33). This transformation is conducted because the statistical properties of unweighted graphs are more amenable to theoretical analysis and because it is easy to define sets of related domains as clusters in the resulting graph (33).

    The transformation described above, however useful, introduces a parameter into the system ($Z_{min}$) and it is unclear *a priori* what value one should choose to construct the graph. The fact that DALI only reports Z-scores of 2 or greater indicates that, at least in this graph, there is an ultimate lower bound for $Z_{min}$, but that point may not be the most instructive for creating the graph. A self-consistent way to determine this parameter involves following the transition in the Giant Component (GC) of the graph (33). In this procedure one steadily increases $Z_{min}$ and builds graphs at cutoffs of ever increasing stringency. Clusters in the graph are defined as a set of nodes for which a path exists in the graph between any pair of nodes in the set; i.e. at each cutoff one determines the disjoint clusters that exist on the graph. The largest cluster so determined is called the Giant Component of the graph. As the cutoff becomes more stringent the GC can only become smaller, and in the case of graphs like the PDUG one can imagine that a transition must occur between a completely connected graph (i.e. a $Z_{min}$ of negative

infinity) and a completely disconnected graph (i.e. a $Z_{min}$ of positive infinity). This transition occurs over a small range of $Z_{min}$ values in the PDUG, and the midpoint of this transition, which occurs at a critical value ($Z_c$), represents an opportune cutoff at which to construct the graph. The PDUG exhibits a $Z_c \sim 9$ (see Figure 3A, taken from (33), for the transition in the GC for the PDUG), and the disjoint clusters on the graph at this point (the Giant Component and all other clusters) correspond fairly well to various levels of taxonomic description in SCOP and CATH (15). Indeed there is some evidence to indicate that the set of transitions in cluster sizes in the PDUG correspond to the natural (human-annotated) levels of structural classification (15).

Given that the distribution of cluster sizes on the graph is the analog of the distribution of fold sizes discussed in the previous section, it is clear that this distribution should follow a power-law, which is indeed the case (6, 33). One of the important features of the graph theoretic representation of the protein universe embodied by the PDUG is that it allows one to start conducting random controls that are difficult if not impossible to imagine in hierarchical taxonomies. One such important control is to simply randomize the entire matrix of Z-scores; that is, to remove all evolutionary information from edges on the graph. Analysis of the resulting randomized graph at the midpoint in its GC transition reveals that the distribution of cluster sizes follows a power law that is nearly *identical* to that observed for the evolved set of relationships (see Figure 3B and 3C, (33)). This surprising result indicates that the cluster (or fold) size distributions that we observe fails the most trivial random control in that they cannot distinguish between the randomized system and the actual PDUG. This result does not indicate that the cluster size distribution in the PDUG is itself a random effect; it could

# Figure 3

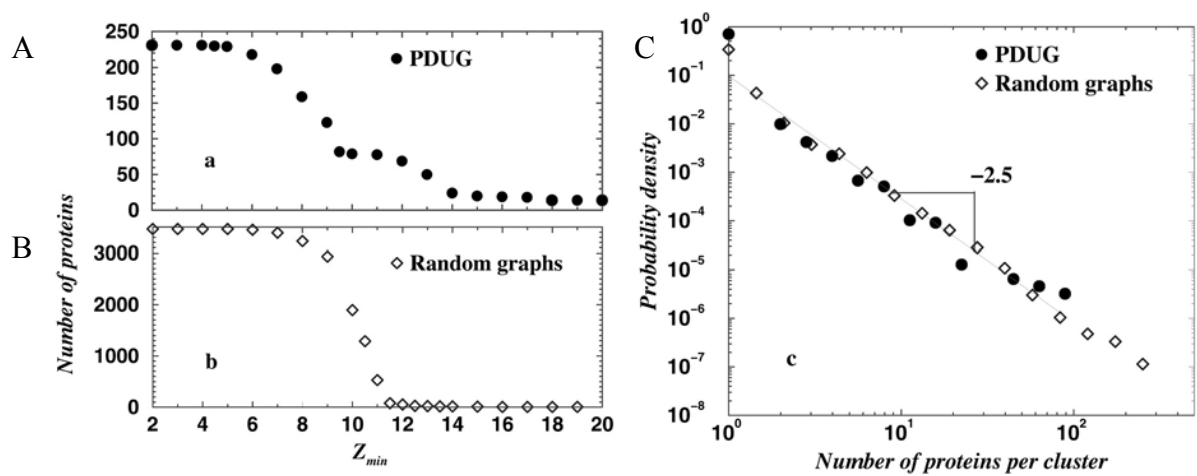

**Figure 3.** Transition in the Giant Component and cluster-size distribution in the PDUG, taken from (33). (A) Transition in the GC for the PDUG; as the $Z_{min}$ is increased, the size of the GC decreases in a fairly sharp fashion with a $Z_c \sim 9$. (B) Transition in the GC for the randomized version of the PDUG. In this case $Z_c \sim 11$. (C) Cluster size distribution for both the PDUG and randomized versions of the PDUG. The x-axis represents the number of proteins in a given cluster, and the y-axis represents the probability of finding a cluster with that number of members. This plot is essentially analogous to the plots in Fig. 2 A and B. Note the striking similarity in the nature of the distributions for the PDUG and the randomized versions.

certainly be a result of specific evolutionary dynamics. It does, however, indicate that cluster size distributions do not contain sufficient information to be useful in testing models of protein structural evolution, despite the many attempts that had been made to do so (6, 26-28, 33). The surprising result of this random control highlights not only the utility of the graph theoretic representation but also the difficulty of interpreting the results of structural analyses in the absence of controls.

The PDUG does differ from its randomized counterparts, however, in terms of its topology. One of the most important topological features of a graph is its degree distribution (34), which represents the probability of finding a node in the graph with a certain number of neighbors. In the PDUG the degree (represented by the variable $k$) of a node is simply the number of structural neighbors it has at that $Z_{min}$, and at the midpoint in the GC transition the distribution of this quantity is well-fit by a power law, i.e. $p(k) \sim k^{-\gamma}$ with $\gamma \sim 1.6$. The degree distribution of the randomized graph is strikingly different and has a Gaussian character at the midpoint of its GC transition, indicating that the degree distribution at least passes this most basic random control (see Figure 4A and 4B, (33)). The distribution of edges per node in the PDUG indicates that this graph belongs to a large class of graphs that have been termed "scale-free networks" (34-39). Although the protein structural universe had been compared to such networks previously (6), the development of the PDUG was the first demonstration that this was indeed the case.

In order to explain the degree distribution, Dokholyan and coworkers developed a divergent, Big Bang (BB) model for protein structural evolution (33). This model is based on the fundamental principle that new domain families, i.e. new nodes on the graph, arise from the duplication of sequences within existing families that subsequently diverge to form new domain families. The rules governing the graphical relationship of nodes to one another are quite simple. Each divergence event is characterized by a random structural distance that is generated to describe the relative similarity between the "daughter" and the "parent" node; in the BB model the distance is chosen from a uniform distribution on the interval from 0 to 1. A cutoff is defined in the system such that two nodes are considered similar only if their distance is below that cutoff. If the daughter node is not similar to the parent node (i.e. if the randomly generated distance is below the cutoff) the newly born node becomes an orphan and is prevented from touching anything else on the graph; the cutoff basically creates a probability for creating nodes that can seed new clusters. If the daughter node *is* similar to the parent an edge between the daughter and parent is placed on the graph. Neighbors of the parent node (the grandmother and siblings of the daughter node) have a probability of connecting the daughter node that is dependent on the distance between the parent and those adjoining nodes (33). Although this model is similar in spirit to the "preferential attachment" models for scale-free networks popularized by Barabasi and coworkers (34, 35), there are many important and clear differences between the two sets of models that allow the BB model to create networks that are scale-free with exponents less than 2 (i.e. $\gamma \sim 1.6$) (33-35, 40). This model is strongly informed by the "free-energy landscape" picture of structural evolution (30, 33), and although it does not model the specific processes or timescales of sequence change at varying levels, the divergent rules that the BB model uses are based on features of this paradigm.

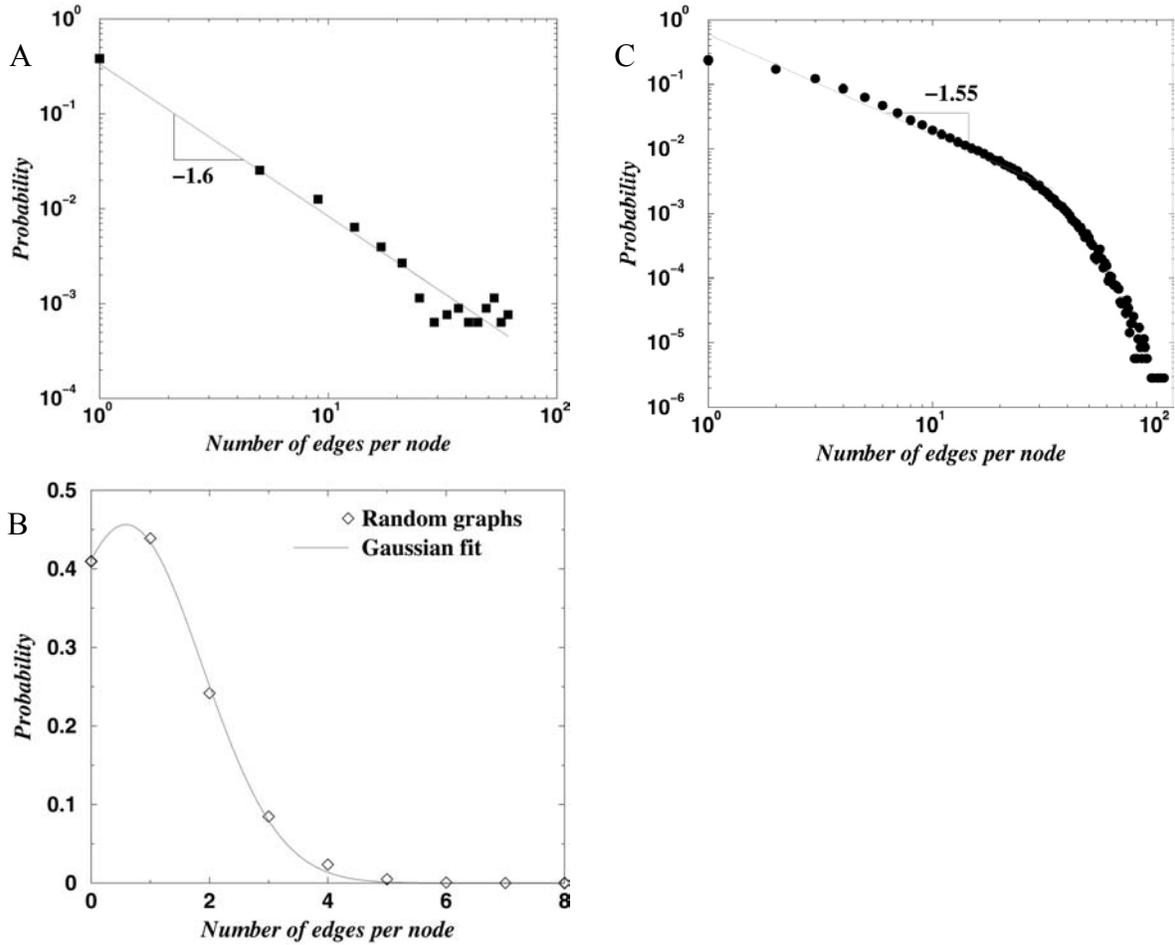

Figure 4. Degree distributions, taken from (33). (A) Degree distribution of the PDUG at a $Z_c$. The straight line indicates a power-law fit with an exponent $\gamma$ of 1.6. (B) Degree distribution of randomized PDUGs at a $Z_c$; note the marked difference between this degree distribution and that of the actual PDUG in A. The solid line indicates a Gaussian fit; the fact that this function well-fits the data supports the interpretation that the randomized PDUGs represent classical random graphs. (C) Degree distribution of an ensemble of graphs produced by the BB model. The straight line indicates a power-law fit with an exponent of 1.55.

Within a certain parameter regime, the BB model produces graphs with degree distributions that are well fit by power-law functions with exponents similar to that observed in the PDUG (see Figure 4C, (33)). As with the earlier models for cluster size distribution, this is quite encouraging for the divergent point of view; it at least implies that divergence is a plausible explanation for the observation of scaling in the structural similarity network. It does not, however, provide conclusive evidence for divergence nor rule out the existence of a competing convergent model that might explain the same graph theoretic observations.

Further exploration of the convergent/divergent debate has taken two separate paths. The first involves understanding the evolution of lattice polymers in a model system meant to replicate the features of protein folding, structure and evolution. The second set of studies involves more detailed analysis of existing sets of protein structures and their distribution in various proteomes. We will first turn to a number of early lessons that were learned from lattice proteins consider the implications they have for understanding the evolution of protein structures.

**Some Lessons from the Lattice**

Much of our theoretical understanding of protein folding, evolution and even function is derived from careful studies of a class of model proteins known collectively as lattice polymers. The monomers or building blocks of these polymers (i.e. the lattice equivalent of the amino acid) are constrained such that they can only occupy a set of sites that make up a lattice in either 2 or 3 dimensions. The monomers are held together by a set of unbreakable bonds of exactly the same length as the lattice spacing. A lattice conformation is defined as a distinct way of "snaking" the polymer through lattice sites without stretching or breaking any of these bonds; basically, such conformations correspond to a self-avoiding random walk on that lattice with a number of steps given by the length of the polymer. Although these lattice polymers represent a very coarse-grained "model" of protein structure and behavior, they present several advantages that make them an ideal system for studying certain features of protein folding and evolution (41-47).

Foremost among these is the fact that all possible conformations of some lattice polymers may be completely enumerated (41, 42, 46, 47). This fact, combined with the limited degrees of freedom that exist for a lattice polymer, has allowed the lattice "folding problem" to be solved both thermodynamically and kinetically. That is, if one defines an alphabet of monomers and a potential energy function governing their interactions, one can determine the energy of a sequence in all conformations and determine whether that sequence will fold into its particular ground state, which may be verified kinetically by Monte Carlo folding simulations (42, 45, 48). Study of such systems has lead to a number of important insights into protein folding, including the discovery that if a sequence exhibits a low energy in a particular conformation when compared with the spectrum of available conformations it will fold into that state (42, 45, 49-51).

Given the extensive work that has been done on folding in lattice systems it is no surprise that a lot of theoretical work on protein evolution has also focused on this class of polymers (43, 46, 47). Many of these studies have actually focused on the nature of sequence distributions and dynamics when one constrains folding to choose a particular structure or small set of structures, a fact that has prevented these systems from providing many insights into the systematics of protein structure discussed at length above. These results are very important in their own right and it is instructive to consider them briefly before turning again to a more structure-centric picture.

One of the most important concepts that arose from the study of model proteins (and of RNA secondary structure) was that of a "neutral network" (47, 52). This network consists of a set of sequences that are related to one another through sets of point mutations that do not influence some critical feature of the protein (i.e. structure of the native state, stability, folding rate, etc.). In very simplified lattice models (2-D square lattices with HP potentials), if one defines a neutral network based on the fact that the constituent sequences all exhibit a particular lattice conformation as their ground energetic state, these networks tend to be organized hierarchically with highly connected

"internal" sequences that are very energetically stable in this native state and less well connected "edge" sequences that are also less stable (47, 53, 54). These results recall the structure of neutral networks in RNA secondary structure, although there are some very important differences between the two cases (55, 56). In RNA, for example, many sequences exist that are in one neutral network but close to the network of another structure; in 2-D model proteins such bridges are much more rare (47, 55, 57, 58). The limited nature of 2-D lattice models and HP potentials notwithstanding, these results provided the earliest glimpse of what the evolutionary landscape of proteins might be like.

More dynamic study of the sequence evolution of 2-D lattice polymers has also been carried out in the context of population genetics (59-61). In these cases structures were considered "functional" and "viable" on the basis of whether or not they could fold into a distinct ground state as measured by a foldability criterion (essentially a folding Z-score, see definition below) (42, 45, 48). Simulations of these systems demonstrated that both designability and evolutionary history have a strong influence on the number and type of structures that are observed. Certain sequences, although highly represented in a completely random search of sequence space, were not as well represented during the population dynamics either due to the topology of the sequence landscape (i.e. structures that have sequences that are not "robust" to mutation) or due to fact that sequence space was not fully sampled during the evolution (59). These results indicate that designability, while it certainly plays a role in the probability of observing a particular structure, may not be the only factor, a conclusion that has been borne out through later work discussed in the section on designability below. Further application of a this methodology provided some evidence that the marginal stability and mutational robustness that are observed for real proteins may result from the nature of sequence dynamics in populations evolving under a folding constraint (60, 61).

One extremely important study of lattice sequences and structures was briefly mentioned above as it provides some evidence for convergence in the convergent/divergent debate. Wingreen and coworkers enumerated all sequences for the lattice 27-mer using the hydrophobic-polar (HP) potential (32). Given the relatively small number of sequences that can be generated using 27 letters in a 2-letter alphabet, they were able to apply each sequence to all of the 103,346 distinct, maximally compact conformations of the 27-mer on the cubic lattice and determine the lowest energy conformation for each sequence (32, 41). Their results demonstrated that some lattice structures were "chosen" as the ground state by many, many more sequences than other structures, indicating that the concept of designability seemed to fit the behavior of these lattice polymers fairly well. Indeed, the distribution of sequence volumes per polymer was somewhat similar to the distribution of fold family sizes and lent some weight to the idea that convergent evolution dominated by designability differences might not be a bad description of actual protein evolution (see Figure 5, (6, 32)). Although a number of caveats about this work may be raised (including the potential problems that HP models have in defining energetically distinct ground states in 3-D lattices), from the results discussed above it is clear that any understanding of protein evolution either on the lattice or off of it must at least consider designability to some extent.

Much of the work that has been done on sequence-based structural evolution in lattice systems has considered only the simplest possible "functional" criterion; namely,

# Figure 5

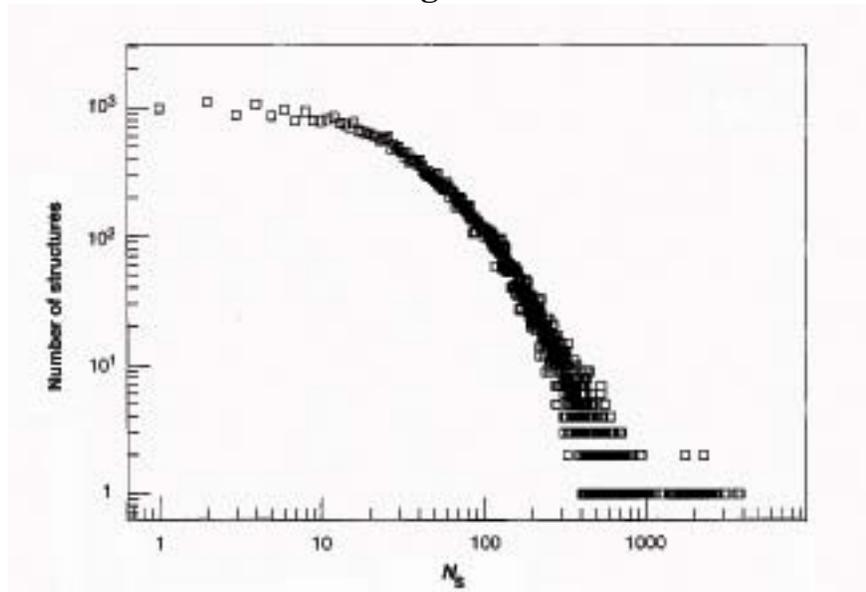

Figure 5. The distribution of sequence occupancy in a complete enumeration of lattice polymer sequences, modified from (32). The x-axis represents the number of sequences that correspond to a given lattice structure and the y-axis represents the number of structures in the analysis that exhibited that number of sequences. Note the similarity between the power-law tail of this distribution and the actual fold size distributions in Figure 2A and 2B and the similarity to the Pareto distributions described in (6)

that of rapid folding to a distinct native state. Attempts have been made, however, to model higher levels of functionality in 2-D and 3-D lattice systems. Most notably, Hirst and coworkers (47, 62-64) have explored the effects of functional constraints on "neutral" networks (at least from the perspective of native state and folding ability) using an HP potential in the square 2-D lattice and diamond 3-D lattice. In this case they allowed for non-compact lattice conformations and looked for sets of sequence/structure pairs that demonstrate "binding pockets" for hydrophobic ligands. In this case adaptive evolution for function consisted of increasing the hydrophobicity of the binding pocket. Although the results of this work may be very sensitive to the assumptions of the model (62-64), it is clear that the inclusion of functional considerations such as ligand binding (and, by extension, catalytic active sites, allosteric regulatory sites, etc.) changes the nature of neutral networks quite considerably. These results are consistent with the observation that not all conservation patterns that define the nature of sequence families are likely to derive from a need for stability or kinetic foldability; in some cases conserved residues or CoC patterns may result from functional considerations (30, 31). Reasonable and intelligent incorporation of functional effects into divergent models of protein evolution either on the lattice or off of it represents a challenge to the modeling of structural evolution that has yet to be fully met (47, 62-64).

**Structural Evolution on the Lattice: Graph Theoretic Approaches**

Although the idea of an underlying "structural space" for polypeptides had been proposed and discussed for many years (6, 65), by the time the PDUG was discovered and analyzed there was very little direct understanding of what features this space might have. Although the scale-free distribution of structural neighbors in the PDUG was clearly not a random effect (33), it was unclear whether or not the space of similarities between any set of compact polymers might actually define a similar network. Not only was such a polymeric control important for understanding the uniqueness and evolutionary implications of scaling in the PDUG; it was also important for quantifying and understanding how a protein structural space might influence the evolution of proteins within that space.

As mentioned above it is currently impossible to understand or characterize the complete space of polypeptide structures. Lattice spaces, however, do not share this limitation and many complete structural sets are available for both 2-D and 3-D lattice systems. In our work on this subject, we chose to explore the structural space of the completely compact 27-mer on the 3X3X3 cubic lattice; although this system lacks many features of real proteins (such as secondary structure elements), it has enough structural diversity to provide potentially interesting results yet the space is small enough to allow for all pairwise structural comparisons to be made (66). Despite the shortcomings of this model system, it provided the necessary compact polymer control against which the PDUG could be compared and allowed for the first complete characterization of a physically realistic structural space.

Given that a hierarchical taxonomy of lattice structures would be difficult to envision or build using human intuition, the lattice structural space was used to create a graph in much the same manner as the PDUG. Each distinct lattice conformation was defined as a node, and all pairwise structural comparisons between nodes were conducted in order to build edges between these nodes. The structural comparisons, in analog to the DALI Z-score, were based on the overlap in contact maps between two structures, and the distribution of the lattices "S-scores" at large values was very similar to the distribution of Dali Z-scores in the PDUG (33, 66). Edges in the Lattice Structure Graph (LSG) were placed when any two nodes have an S-score greater than the cutoff $S_{min}$. Degree distributions for all structures in the LSG at each value of the $S_{min}$ cutoff were well-fit by Gaussian functions and are much more akin to classic random graphs (and the randomized PDUG control) than the scale-free PDUG characteristic of structural evolution (see Figure 6A, (33, 66)). It is thus clear that a scale-free network of structural relationships is at least not a *generic* feature of compact polymer spaces.

**Figure 6**

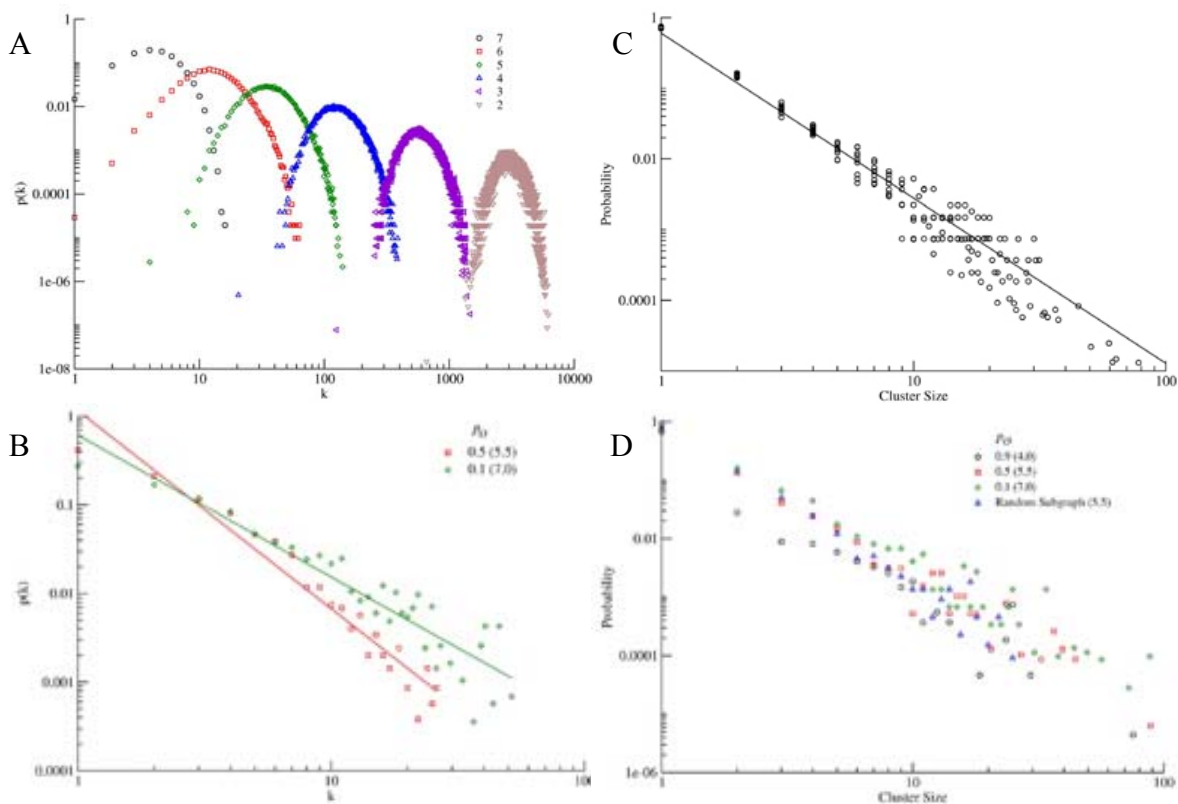

**Figure 6.** Results from the analysis of the LSG, taken from (66). (A) Degree distributions of the entire LSG at a variety of $S_{min}$ values as indicated. All of these degree distributions are well-fit by Gaussian functions and lack the characteristic power-law features found in the PDUG. (B) The degree distributions of two subgraphs sampled from the LSG using the divergent evolutionary algorithm described in the text at 2 different values of the orphan-creation probability ($p_O$). Both are well-fit by power-law functions (solid lines) and the data set indicated in green exhibits a power-law exponent $\gamma$ of 1.6. (C) Cluster size distributions for a number of random subgraphs of the LSG. The cluster size distributions of these random subgraphs are virtually indistinguishable from the cluster size distributions of the PDUG, randomized PDUG or any other type of subgraph of the LSG. The solid line indicates a power-law fit of the distribution from a representative random subgraph. All of the graphs were analyzed at the midpoint in the transition of the GC. (D) A comparison of the cluster size distributions of evolved subgraphs (with degree distributions from B) to the random subgraphs from C.

The random-graph-like degree distribution in the LSG, although it is not necessarily indicative of the nature of the space of actual compact polypeptides, nonetheless suggests the following question: if an underlying space of structural possibilities is not scale-free, is there any way in which the space might be sampled in order to produce scale-free networks? This view of structural evolution poses the problem not as one of domain "creation" or "innovation" but rather as a sampling process through which structures that exist in the space of possibilities come into existence when a sequence is found that can fold into them. As one would expect, sampling 3500 structures (i.e. a number of nodes equivalent to that in the PDUG) from the LSG completely randomly (i.e. choosing structures with equal probability from the graph) does not produce scale-free networks (66). Some sort of sampling bias is required to obtain scale-free subgraphs of the LSG.

One might first imagine that one could choose lattice structures on the basis of their designability. As we will discuss in much greater detail below, for lattice structures it is possible to accurately estimate the designability of a conformation based on features of its contact map (67), and one can imagine that evolution might sample only those structures that are highly designable. The resulting set of structures does not, however, represent a scale-free network, although it does differ from a random subgraph of the LSG (66). The main difference between the random subgraph and the set of highly designable structures is the average connectivity—highly designable structures tend to share structural features and so the average number of connections for that subset at a certain cutoff is much higher than the average number of connections for the random subset. Although this result does not guarantee that sampling according to designability in a real polypeptide space would not yield a scale-free network of structural similarity, doing so in the LSG is clearly insufficient to obtain scaling.

Another set of sampling algorithms involves dynamically choosing structures from the graph in accordance with evolutionary rules. Such models are similar in spirit to the original BB model but, rather than creating nodes out of thin air and ordering their similarity according to an (ultimately) arbitrary set of rules, they are constrained to choose only existing lattice structures as nodes and must abide by the existing structural relationships that constitute the LSG. Such algorithms may utilize the structural relationships in order to choose nodes, but they cannot set or change those relationships. We found that certain classes of models based on "duplication and divergence" could indeed sample PDUG-like scale-free networks of structures from the LSG (see Figure 6B, (66)). Duplication events in these models involve picking a random structure from the evolved subset and choosing a daughter structure with some particular structural distance from the parent (which is equivalent to the structural divergence that occurs during duplication in the original BB model). These models differ strongly from the original BB model in the manner in which orphans are chosen, however. In the original algorithm, structures that diverge beyond a cutoff do not contact the parent node and are prevented from contacting any node that currently exists on the graph. In the lattice model, however, simply choosing some daughter nodes that are beyond a given structural similarity cutoff did not yield PDUG-like behavior; in essence, even structures that are very far from any given parent have an appreciable probability of being similar to *some other* structure in the evolved subgraph and thus do not necessarily represent "true orphans" (33, 66). In order to evolve scale-free networks it was necessary to introduce a probability that a daughter would represent a true orphan and choose such orphans from the ensemble of structures in the LSG that did not exhibit strong structural similarity to any evolved structure at that particular point in time. The inclusion of this global similarity calculation represents a marked departure from the original, purely local, BB model and poses the interesting question of how sequence evolution alone could choose such orphans (a point discussed in greater detail below).

At this point it is important to stress the necessity of using specific topological observables (such as the degree distribution) to compare models with one another, the LSG and the existing protein universe. At the transition in the GC, every type of LSG subgraph exhibits a scale-free distribution of cluster sizes with exponents similar to one another and to that of real proteins (see Figure 6C and 6D, (66)). It is thus impossible in this system to identify a graph as having been sampled randomly, according to designability or according to divergent evolutionary rules simply on the basis of the cluster size distribution. The differences in degree distribution between the different sampling methods are, however, extremely clear. This result was foreshadowed by the randomized PDUG control but it is nonetheless a clear example of the limitations of considering only cluster size distributions (6, 26-28, 33).

Although the results outlined above are important as a proof of evolutionary principle, they do not represent a clear bridge between the sequence-based studies of lattice structures outlined above and the topological structure of the PDUG. The application of the BB model to the LSG has a clearly dynamic and divergent character, however, it is impossible to know whether the "divergent" rules discussed in the preceding paragraphs accurately represent the types of structural changes that occur during sequence divergence. It is fairly obvious that organisms cannot perform the kind of structural comparisons required by the algorithm as they evolve; for instance, an organism cannot explicitly determine whether or not a newly evolved domain is a structural orphan (66). One of the important questions raised by the LSG work involved understanding if sequence dynamics alone could provide the basis for scale-free networks of lattice structures; that is, whether it was possible to bridge the gap between the work of Taverna and Goldstein and the graph-theoretic approach to understanding structural relationships and diversity (33, 59, 66).

To approach this problem we built a completely physical, sequence-based model of structural evolution that relies on absolutely no structural information to build a subgraph (68). This model is based on the folding of 27-mer sequences into compact lattice structures and is based on a specific alphabet and potential energy function for which folding can be evaluated (which in our case was the Mirny-Shakhnovich potential in all 20 amino acids) (69-71). Running Monte Carlo folding simulations of lattice polymers, while certainly not nearly as computationally intensive as folding simulations of proteins with realistic degrees of freedom, is nonetheless computationally costly and so rather than verify folding directly we employed an energy Z-score (here called an F-score):

$$F_{i,k} = \frac{E_{i,k} - \langle E_i \rangle}{\sigma_{E_i}}, \tag{1}$$

where $F_{i,k}$ is the F-score of sequence *i* in conformation *k*, $\langle E_i \rangle$ is the average energy of sequence *i* in all conformations, and $\sigma_{Ei}$ is the standard deviation of the energy of sequence *i* across all conformations. As mentioned briefly above, this measure of foldability is derived from early work on protein folding from the spin-glass perspective (45, 49, 50). A sufficiently low value of this F-score for a given sequence in a given lattice structure indicates that the sequence is much more stable in that structure than it is in the "non-native" ensemble and implies that the sequence will fold into that structure. This principle has been used many times to design sequences to fold into lattice structures quickly and with high stability (48, 59-61, 68).

The evolutionary dynamics of this model are quite simple and are directly based on the motivation behind the original Big Bang model and the LSG-based model discussed above. At first, one chooses a "seed" structure at random from the set of all possible structures and "designs" an amino acid sequence to fold into this structure. At each step in the simulation an existing sequence is chosen from the evolved set at random and is duplicated. A number of mutations are then made to this sequence during the divergence step (in our case the number of mutations is set to be the same for all duplication events). The "native state" of the new sequence is determined by finding the compact lattice conformation in which the sequence exhibits the lowest energy and folding is assayed using the F-score. If the F-score of the sequence in its native state is below some cutoff, folding is assumed and the sequence-structure pair is added to the graph in (basically) exact analogy to the schema employed by Taverna and Goldstein in their earlier work (59, 68).

If one sets the F-score to positive infinity, that is, if one does not specify an F-score cutoff but accepts every new sequence along with its lowest-energy conformation, it is possible to obtain scale-free LSG subgraphs with PDUG-like exponents (Figure 7A). The level of sequence divergence that is required to obtain these results is relatively small—only 2 mutations are made per duplication step. When there is no folding

# Figure 7

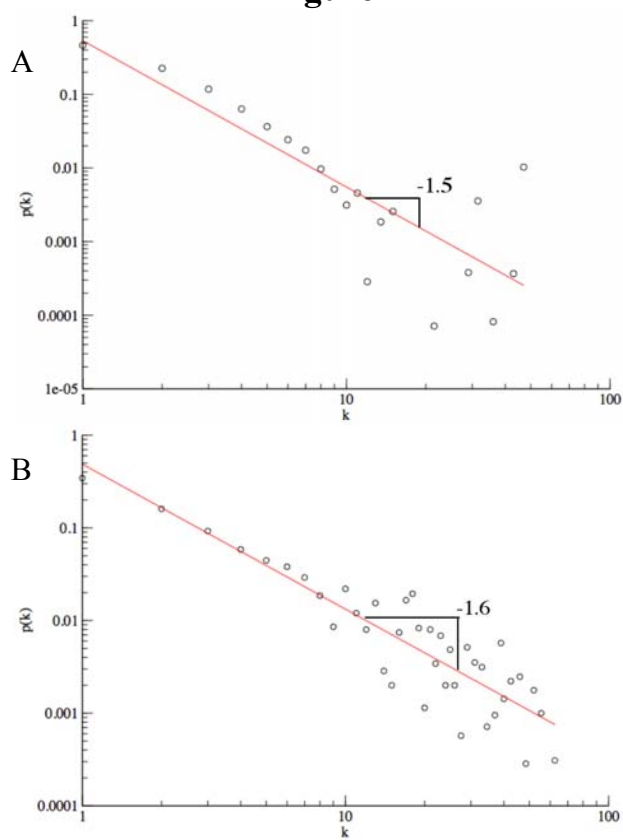

**Figure 7.** Degree distributions from sequence-based models of protein evolution on the lattice, taken from (68). (A) Degree distribution for an LSG subgraph based on the sequence-based evolutionary algorithm without a folding constraint. The solid line represents a power-law fit with an exponent $\gamma$ of 1.5, close to that observed in the PDUG. (B) Degree distribution of a graph evolved using a folding cutoff. The solid line corresponds to a power-law fit with an exponent of 1.6, close to that observed in the PDUG.

constraint the algorithm generates considerable structural diversity in a small sequence volume, demonstrating that many different types of "native" structures live within a small sequence neighborhood of a given starting sequence-structure pair (68). Introduction of a folding cutoff, however, drastically increases the level of sequence divergence that is required to observe PDUG-like behavior. This is to be expected; although many sequences that are close to a starting sequence may have native states somewhat different from the starting structure, most of the cases in which those sequences are *stable* in their respective native state involve structures that are very similar (if not identical) to the starting structure. Thus 8-10 mutations per step are required to observe scale-free networks with the right range of exponents when the F-score cutoff is set to –6 (see Figure 7B, (68)).

The folding cutoff also influences other features of the model's behavior. For one, the model starts to exhibit "non-ergodic" characteristics in that different starting sequence-structure pairs lead to differences in the nature of the evolved subgraphs. For instance, one randomly chosen seed structure produces PDUG-like degree distributions when the number of mutations is equal to 8; another only does so when the number of mutations is set to 10. Even different simulations with the same initial condition do not converge to graphs of similar behavior over the course of 3500 steps. Neither observation holds true in the absence of the folding cutoff, indicating that the folding constraint may make the evolutionary landscape considerably more "rugged" (68), a finding that is in complete agreement with earlier work on the nature of evolutionary landscapes (47, 72). Although the sources of this ruggedness are not currently well understood, it may have to do with the designability of structures in the regions of the evolutionary landscape surrounding a given starting structure. Those simulations in which a very designable structure is discovered early may have trouble developing the level of structural diversity observed when such a structure is discovered relatively late or not at all. It is well-known that early events in some models for scale-free networks can cause "giant fluctuations" in the nature of the graphs they produce (73), but much further work remains if we are to understand why this occurs in these models and what implications these findings may have for the evolution of real protein structures.

Regardless, the work outlined above has three major implications. First, the degree distribution of the PDUG is not a trivial result of the fact that it is constructed from a set of compact polymers. Secondly, if the underlying space of polypeptide structures shares topological features with the LSG it is possible to define divergent evolutionary rules that sample its random-graph topology in such a way that scale-free networks may be produced. Finally, it is possible to define a model based only on the divergence of actual sequences that fold into lattice structures that recapitulates the PDUG's graphical features. This last point is an important proof that the divergent principle, when actually modeled mechanistically, has the right ingredients to explain the protein universe as we observe it today (33, 66, 68).

**Results from Structural Proteomes**

Studies on the lattice, however satisfying and important, are ultimately limited in the amount of information they can provide regarding the evolution and natural history of protein sequences. As mentioned above the plethora of plausible and successful divergent models that exist to explain the statistical properties of the protein universe cannot really be taken to prove that convergent models cannot provide similar explanations. In order to garner evidence against the convergent picture (rather than more evidence for divergence) it is necessary to turn back to the realm of actual protein structures and consider how those structures distribute themselves within the proteomes of existing organisms.

In the spirit of developing a convergent picture of structural evolution, let us consider a simple thought experiment. Suppose for a moment that the underlying space of polypeptide structures had a distribution of structural similarity that was scale-free and somewhat similar to that observed in the PDUG. Although this is certainly not true of the LSG (see above), there is currently no way to actually demonstrate that this is not the case for polypeptide structures with realistic degrees of freedom. Now suppose that sequences sampled this space of structures randomly and completely independently—that is, consider the "null" case of a convergent model. In this case one would observe a PDUG-like degree distribution (74) without any need for divergent mechanisms. Thus, despite the promising success of divergent models in recapitulating the degree distribution of the PDUG, it is surprisingly easy to create a model that explains this phenomenon from a simplistic convergent perspective.

In a sense, the only way to make strong, well-supported statements in opposition to either the convergent or divergent paradigm is to test both of them against some observation and see if one does not fit the observed data. Although the PDUG can provide a large amount of evolutionary information, it is a single, static dataset, and trying to extract dynamic information from it is akin to trying to reconstruct the trajectory of an object based only on a snapshot of its final position. In the case of structural information, however, the existence of structural proteomes serves as a potential source of dynamic information. Given that organisms have been speciating from one another for billions of years and given that speciation events largely separate proteomes from one another (although the separation is not necessarily complete (75-78)), it is clear that the distribution of structures in organisms may actually provide important observables against which both convergent and divergent models may be tested.

The importance of genome-specific sets of structure, or "structural proteomes," to the study of protein evolution has long been recognized. As the levels of protein systematics were developed, researchers immediately began wondering how they were distributed between and within the three kingdoms of life. Koonin and coworkers discovered that, although there are certainly many differences between the different kingdoms of life in terms of their fold content, there is also a great degree of similarity, a fact that has also been shown to hold true in comparisons of the large eukaryotic clades (79, 80). These results imply, at least on a superficial level, that structural proteomes might contain important "dynamic" information that cannot be obtained from the PDUG. Constructing organismal PDUGs (oPDUGs) is fairly straightforward. A node in the PDUG corresponds to a family of related sequences, and so if one finds at least one member of that family in the proteome of a given organism, then that domain family exists in that organism (74, 81). This procedure allows for the extension of the solved structures represented in the PDUG to a large number of organisms despite the fact that the structure of a particular domain may not have been solved in all (or any) of the organisms in question. For those organisms with fully sequenced genomes one may create as complete a structural picture as possible by determining the domain compliment of that proteome. Each organismal list of domains represents a specific subgraph of the PDUG, and analysis of these oPDUGs has the potential to reveal dynamic information regarding the evolution of the protein universe. In the case of prokaryotic (bacterial and archaeal) proteomes, we demonstrated that the oPDUGs exhibit scale-free networks with exponents that are, on average, close to the exponent observed in the PDUG (74).

If one considers the null convergent model above, it is clear that organisms evolving under this type of model will exhibit a specific type of subgraph. Starting from a least universal common ancestor, the evolved structural universe accrues structures at random from the underlying space. Subsequently, speciation occurs, resulting in two proteomes that randomly sample this space independently. After billions of years of evolution and many speciation events, the sum total of all (randomly chosen) structures in existence across all organisms will clearly constitute a random subgraph of the overall structural universe and, if the underlying structural space represents a scale-free network, the degree distribution of observed in the PDUG is fairly easy to explain. This model also predicts that each independent proteome will constitute a random subgraph both of the underlying structural space and of the PDUG (74).

This null hypothesis of convergence was easily tested. We considered the oPDUG of an organism with a fully sequenced proteome (say *Bacillus subtilis*) and created random subgraphs of the PDUG that contain the same number of nodes as the oPDUG in question. As one might expect, these random subgraphs exhibited scale-free degree distributions similar to the *B. subtilis* network in terms of their exponents. These random subgraphs were strikingly different from the evolved oPDUG, however, in terms of the maximal connectivity found in the graph ($k_{max}$). Indeed, for many proteomes we found that the $k_{max}$ of the evolved graph is much, much greater than the random counterpart, and this discovery indicates that the convergent null hypothesis is not likely to be true. Using basic graph theoretic approaches it is actually possible to quantify the probability that a particular oPDUG is a random subgraph of the PDUG (74), and in the case of over 2/3 of the prokaryotic genomes considered, the P-value of randomness was less than $10^{-6}$ (see Figure 8A). This finding indicates that, although the null convergent model is fully capable of explaining the degree distribution of the PDUG it cannot explain the observed topological features of individual structural proteomes.

Although the null convergent model is fully consistent with the convergent picture of structural evolution, it is perhaps not surprising that organisms do not contain a random compliment of structural domains—it is difficult to imagine, for instance, that a fully functional set of biochemical machinery could be based on such a sampling method. One might imagine that either functional or organismal constraints could lead the system to sample structures in a biased, but nonetheless convergent, manner. It is clear that this bias must in some way correspond to an increased probability of choosing domains with higher values of $k$ in order to overcome the lower $k_{max}$ in unbiased random subgraphs. The bias that is necessary to overcome the decrease in probability at high $k_{max}$ is so strong, however, that it creates unrealistic oPDUG degree distributions and as such it is not a likely explanation for the oPDUGs we observed (74). The results outlined above indicate that a whole class of convergent models in which the state of the organismal graph at time $t_0$ cannot influence the sampling of domains from the underlying graph at any later time $t$ (a class of convergent models that we call "equilibrium convergence") is unlikely.

It is quite easy to modify the BB model of structural evolution to include a mechanism of speciation. In this case structural evolution proceeds divergently exactly as in the BB model. The only difference is that evolution begins in a single proteome.

# Figure 8

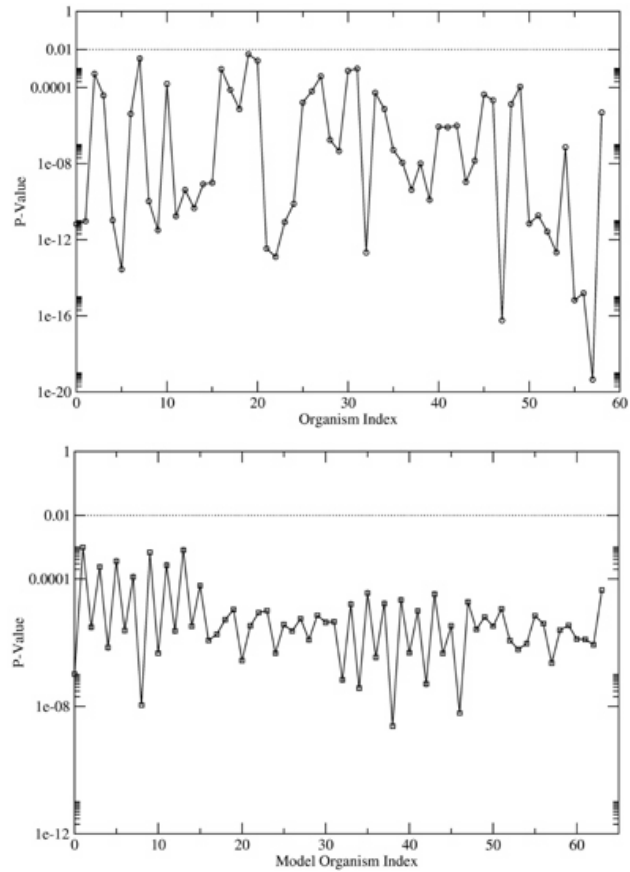

**Figure 8.** Random subgraph P-values, taken from (74). (A) P-values for actual oPDUGs. The Organism Index is an arbitrary number chosen to represent each oPDUG. The dotted line indicates a 1% chance that the organismal subgraph is a random subgraph of the PDUG. (B) A graph as in A but for the BB speciation model. P-values in this case represent the probability that a particular model organismal subgraph is a random subgraph of the overall model PDUG produced by the BB algorithm. The dotted line indicates a 1% chance that the subgraph represents a random subgraph.

After time this proteome undergoes a speciation event, producing two identical lineages. These lineages evolve independently from one another with the restriction that any given duplication event must occur *within* a single proteome; that is, at each step a parent domain can only give rise to a daughter domain in the proteome in which the duplication occurred. The model oPDUGs produced by this algorithm all exhibited much larger values of $k_{max}$ than random subgraphs of the resulting model PDUG and, as such, had very low P-values of being random subgraphs themselves (see Figure 8B, (74)). It is thus clear that, while a whole class of convergent models cannot explain the oPDUGs observed in nature, a divergent model easily can.

The results discussed above do not indicate that convergence has never or can never occur, and the debate about the nature of protein structural evolution is still proceeding. Although many researchers now agree that the divergent scenario is more likely (6), the only existing direct evidence *against* any convergent model(s) has been derived from careful analysis of the oPDUGs (74). Further study of these and other graph theoretic features of the protein structural universe are necessary to explore the implications of this observation in further detail and understand the extent to which structural convergence of different types and on different scales may have occurred over the course of structural evolution (82, 83).

**Frontiers**

Despite the progress that has been made in the past several decades towards understanding the evolution of protein structure, it is clear that the picture we have is still a long way from complete. One of the most important efforts that are currently ongoing in this field is structural genomics (6, 84, 85). Although our structural coverage of some types of proteins and of some organisms is not bad (6, 24, 79, 80), data in other areas of the organismal and structural universe is sparse and, as more and more structures are solved, our ability to ask and answer meaningful questions regarding structural evolution will increase. Structural genomics, aside from simply providing more data amenable to evolutionary analysis, should also give us the most unbiased view of a number of structural proteomes to this date. This data is crucial to our efforts to build ever more accurate models of structural evolution and ever more complete pictures of the natural history of the evolved protein universe.

The structural evolution models discussed above, for all their success, do have a number of failings that present important challenges to the development of a global understanding of structural evolution. One such failing is the fact that none of these models accurately predict the number of orphans that one should observe in the PDUG. Indeed, these models often predict fewer orphans in the graph than expected from the power-law fits of their degree distributions; the PDUG, however, exhibits far more orphans than predicted by its power-law fit (33, 66, 68, 74). It is very encouraging that the sequence-based model of structural evolution on the lattice predicts a predominance of orphans given that this model (and this model alone) contains no mechanisms geared specifically at introducing orphans into the graph (68). Despite this success the number of orphans is still too low, and it is clear that ensuing models of structural evolution will have to deal with this deficiency. Given that certain orphans represent *truly* new topologies, they may (on average) have the ability to access more novel functions than non-orphans. If we imagine that duplication and divergence are occurring frequently in most populations, structural orphans could have a higher probability of being retained by the population compared to non-orphans simply because they have a higher probability of offering a selective advantage to the subpopulation in which they occur. This would represent an effective increase in the creation of orphans and might present a very intriguing link between these types of structural evolution models and the selectionist viewpoint favored by some (80).

Many interesting questions remain regarding the sequence-structure landscape for protein evolution in lattice systems. One area of intense interest is the influence of designability on sequence and structure dynamics. Although this has been explored to some extent in the text above and is discussed further in the next section, integrating the designability picture with our understanding of structural diversity and evolutionary change is an important challenge that may be best approached by further study of lattice models. In the future, efforts to map the size, shape and connections between "neutral networks" in sequence space could provide a number of clues not only as to the specific influences of designability but also to the nature of sequence transitions that drive structural transitions. An explicit map of the sequence/structure landscape should provide a much deeper understanding of the "free-energy" picture that motivates the current models and understanding of structural evolution (30).

At this stage it is important to note that, despite the success of various stochastic models of protein evolution, the fact that all of these models lack any functional description or considerations has lead to the proposal that these models are fundamentally inaccurate (80). While one may certainly argue that the BB models are incomplete (they are each, after all, very coarse-grained descriptions of biological and/or structural reality), this does not imply that their correspondence to observed topological features in the PDUG is necessarily serendipitous. Indeed, these models indicate that duplication and divergence is sufficient to obtain PDUG-like behavior, which represents a supremely important proof of evolutionary principle. Designability and functional utility may certainly impact *which* folds are highly populated and *which* domains have high degrees—this point is explored further in the next section. Regardless, such considerations are not necessary to describe the overall statistical properties of the protein universe, and this implies that these properties are most likely a fundamental consequence of the underlying mechanism of duplication and divergence upon which evolutionary selection may act. BB-type models imply that both selection and duplication/divergence may affect the distribution of protein structures, and features of these models provide some explanation of the uneven distribution of populations even among families and folds that perform the same function (6, 86).

Although the convergent/divergent debate is considered dead by some, there are many aspects of structural evolution that may be informed by a convergent picture. As mentioned above, a divergent picture of structural evolution is certainly in no way incompatible with the "convergent" concept of designability. It is also clear that functional constraints may actually cause the convergence of local structural elements in structures that clearly belong to different folds; that is, structural convergence may occur in the active sites of certain enzymes that are clearly unrelated at the cluster/fold level of topological description (6, 68, 74, 82, 83). Much like the physical nature of flight constrains the nature of a wing and thus leads to convergent similarity in independently evolved wing structures, so the physical reality of catalysis may cause "convergent" positioning of residues, functional groups or even elements of secondary structure within a small region of a domain. This proposal is in no way inconsistent with the divergent picture of domain evolution; in this case local similarity can easily arise in the context of global structural expansion.

## Designability and Protein Evolution

**Structural Determinants**

Although the concept of designability has been central to the convergent/divergent debate, especially on the convergent side, the concept does not depend on a convergent picture of structural evolution. Although the motivation for proposing this concept, i.e. the observation of heterogeneity in fold space (6, 16-18), may not ultimately derive solely from differences in designability, it is clear that, at least in principle, some types of protein structures may correspond to a greater number of sequences than other structures. Many researchers have demonstrated that lattice polymer systems exhibit certain structures have much larger designabilities than others (32, 59). Given the very complete nature of many of these studies, the designability effect seems to be quite real, structure-specific, and independent of any specific evolutionary mechanism. Despite this clear demonstration of the concept, these studies did not provide many quantitative insights as to the specific features that make some protein structures more designable than others.

Early analysis of real and model protein structures that seemed to have high designability (such as the TIM barrel) indicated that certain structural features such as symmetry defined these types of structures (32, 87). This was followed by work that capitalized on this idea and attempted to develop a structural predictor of designability with somewhat limited success (87, 88). Recently, England and Shakhnovich developed an analytical approach that served as the basis for a designability measure (67). They demonstrated that (under certain assumptions) the per-monomer free energy ($f$) of a structure in sequence space could be expressed as a function of traces of powers of the contact map, i.e.:

$$f = -\frac{1}{N} \sum_{n=2}^{\infty} \text{Tr}(\mathbf{C}^n) a_n, \qquad (2)$$

where $N$ is the number of residues in the polymer, $\mathbf{C}$ is the residue-residue contact matrix defining the structure, and $a_n$ is a positive weight that depends on the residue-residue contact energies from a potential energy matrix $\mathbf{B}$. The trace of powers of the contact map sounds like a very abstract quantity, but it actually has a fairly clear and intuitive physical interpretation. The trace of the $n^{th}$ power of a contact matrix represents the number of $n$-member cycles in that matrix, which indicates that the contacts between these $n$ amino acids are ideally arranged to provide maximal stability, thus relaxing the stability constraints on the rest of the sequence (67). Although this measure is based on certain assumptions (such as the idea that protein energetics may be completely represented through a pairwise contact potential and some assumptions about the nature of the potential energy function $\mathbf{B}$), it is a very good predictor of designabilities measured directly for lattice proteins where the energetics are explicitly constrained to a contact form (see Figure 9, (67)). Even though the designability measure was not used to evaluate a complete sample of either sequence or lattice structure space as performed by Wingreen and coworkers (32), it is at least reasonable to assume that many of the structural influences on designability of lattice structures are being captured by this measure.

The England-Shakhnovich (ES) designability measure introduced above has been employed in a number of subsequent studies, including the LSG work discussed in the preceding section. In the first demonstration that this measure might capture some

# Figure 9

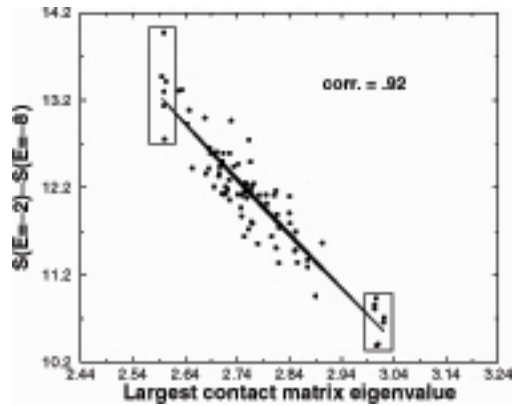

Figure 9. Success of the ES designability measure for lattice proteins, taken from (67). In this plot the x-axis represents the ES measure of designability as outlined in equation 2. The y-axis represents empirically determined sequence entropies from simulations in which lattice sequences were designed to fold into specific lattice structures (i.e., tent to exhibit low F-scores as discussed previously). The correlation between the predicted and observed entropy in this case is quite good (R = 0.92), indicating that the ES measure is a successful indicator of designability at least for lattice proteins.

important features of protein stability and designability off of the lattice, England, Shakhnovich and Shakhnovich showed that thermophilic proteins are statistically significantly more designable than mesophilic proteins (89). Given that the theory behind the ES designability measure in Equation 2 predicts that more designable structures will exhibit many more stable sequences than less designable structures, this finding lends some weight to the claim that ES has some bearing on the designability of real proteins (67, 89). Although it is clear that there is no consensus in the field regarding the applicability of Equation 2 to existing protein structures (47, 67, 89), the considerable success of this measure for lattice proteins combined with burgeoning evidence from existing protein structures at least provides some evidence that this geometric measurement has the potential to capture much of the relevant designability signals from protein structure.

**Influence of Designability on Structural Evolution**

In order to understand the combined influence of designability and divergence on the evolution of protein sequences, Tiana and coworkers developed a 3-D lattice simulation similar to the sequence-based divergent lattice algorithm discussed above (59, 68, 90). In this case a seed sequence/structure pair is chosen to start the simulation, and duplication and divergence are employed to not only increase structural diversity but also simulate the creation of paralogs and orthologs. This is achieved by allowing duplication and divergence that is constrained to a single structure (i.e. mutations are only accepted if the same native state is chosen by the sequence) in addition to allowing mutations that change the structure. The authors found that the structures produced by these divergent simulations resulted in a striking increase in the designability over time. This increase was measured using the trace of the $8^{th}$ power of the contact matrix (the $Tr(\mathbf{C}^8)$ term in Equation 2) and by direct verification of designability through sequence enumeration (90).

This very interesting observation implies that sequences starting from structures with a given designability tend to "migrate" towards structures that are more designable. Given the very nature of designability (i.e. the volume of sequence space corresponding to a given structure) this is perhaps not surprising: the simulation is simply more likely to find stable, folding sequences for structures that are more designable. This is nonetheless an important observation that builds on the work of Taverna and Goldstein (59) in that it implies that both designability and evolutionary dynamics may have a role to play in explaining protein evolution. Although the authors did not consider the topological features of the structural space they generated in that particular piece of work (33, 66, 68, 90), it is clearly possible to apply the graph theoretic description of the PDUG and LSG and its associated BB model to asses whether these simulations with increasing designability over time lead to model structural universes similar to the existing one.

To test the prediction that designability (as measured by the $\text{Tr}(\mathbf{C}^8)$ parameter) is increasing over time in real proteins, the authors compared the designability of all eukaryotic-only domains from the PDUG to prokaryotic-only domains. They found a significant difference in the designability distribution: prokaryotic domains are, on average, less designable than eukaryotic domains. From this they claimed that, since prokaryotic organisms are less complex and "evolved before" eukaryotic organisms, designability has been increasing over time for the real protein universe (90). It is impossible, however, to argue that the prokaryotic organisms are *a priori* more structurally basal than the eukaryotes simply based on the fact that the prokaryotic cellular organization most likely evolved before the eukaryotic one. Indeed, if we assume that a Least Universal Common Ancestor (LUCA) exists for eukaryotes and prokaryotes, by definition the proteins specific to each lineage have been evolving separately for *exactly* the same amount of time. Since there is no reason to believe that structural innovation has proceeded more rapidly or has occurred on balance more recently in the eukaryotic lineage it is difficult to claim that the difference in designability the authors observed (90) is actually due to a tendency for proteins to increase in designability over time. More careful phylogenetic analysis (as discussed in the next section) may shed some light on this problem. This is not to say that the analysis is necessarily incorrect; it is simply impossible to judge without much more careful study.

Despite the limitations of the analysis mentioned above, further application of the ES designability to real protein sequences and structures has provided clear evidence that designability has had at least measurable influence on sequence and structural evolution. Shakhnovich, DeLisi and ourselves (91) recently demonstrated that the ES measure of designability (calculated in this case using the simply the contact density, or CD, based on the first-order term in Equation 2) correlates very well ($R^2 = 0.9$) with the logarithm of the observed size of sequence families (Figure 10A). In this case families are defined in exactly the same way that nodes on the PDUG are defined: that is, a set of related sequences all of which (presumably) fold into very similar structures (i.e., structures with essentially the same contact maps and thus the same values of $\text{Tr}(\mathbf{C}^n)$). Considering the structural neighborhood of a given node (i.e. averaging over a set of neighbors on the PDUG) produces a similar correlation (Figurre 10B). These results indicate that, at least *on average*, the geometric measure of designability represented by CD is a fairly good predictor of family size. CD is also a good measure of the functional flexibility of a domain: the higher the CD, the more functions performed (on average) by the family of sequences.

The correlations mentioned above are calculated based on averages within sets of domains that belong to either CD or family-size bins and are thus statistical in nature. When one looks at the raw correlation (without binning), the correlation drops to an $R^2$ of 0.09 (Figure 10C). This correlation is still very statistically significant, but it is clear that, for an individual family, CD is not a terribly good predictor of family size. Assuming that CD is itself a good measure of designability (as the strong correlation observed for the averages might suggest), the question then turns to what other factors might be influencing family size. The results of both Taverna and Goldstein and Tiana and colleagues (59, 90) suggest that evolutionary history might also play a role. To investigate this somewhat further, we calculated the correlation between (log) family size and CD for the most ancient set of domains as defined by conservation across all three kingdoms of life. In this case, if a domain could be assigned to any one oPDUG from each of the three kingdoms (eukarya, bacteria *and* archaea), then it is phylogenetically reasonable to assume that this particular domain was present in the Least Universal Common Ancestor of all living organisms. These domains demonstrated statistically significantly higher CD values when compared to the entire set of domains in the PDUG, a feature that might have resulted from the thermophilic lifestyle of the LUCA organism (89, 91). The family sizes of these domains were also statistically significantly larger, and the overall correlation between CD and family size within this set improves to an $R^2$ of 0.16 (Figure 10D). These results imply that both designability and evolutionary history impact family size: high designability denotes the *potential* for a large family but natural history determines the extent to which that potential is explored (59, 90, 91).

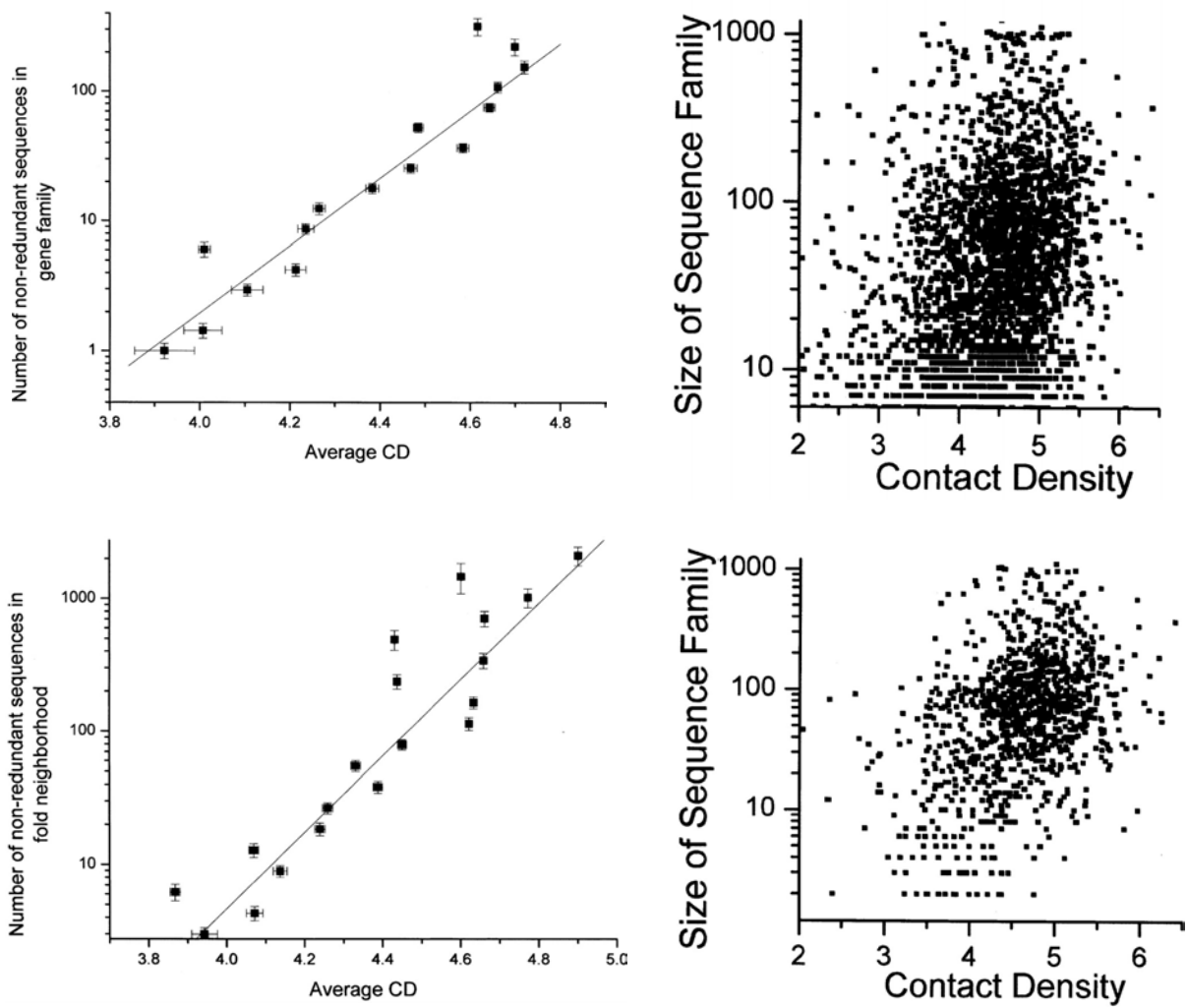

**Figure 10.** Influence of designability and evolutionary history on the size of folds and families, taken from (91). (A) The correlation between the log of the average family size for a given domain and the average CD of domain families within a certain bin in CD. The correlation in this case is very high ($R^2$ of 0.9). (B) A plot similar to that in A, but in this case averages are taken over a structural neighborhood in the PDUG. The correlation here is quite strong as well ($R^2$ of 0.9). (C) Raw correlation of the CD of a protein domain with the size of the particular domain family. Note the much weaker correlation ($R^2$ of 0.09). (D) Raw correlation of familiy size with CD for only those protein domains that have a very high probability of existing in the LUCA of all living organisms. In this case the raw correlation improves to an $R^2$ of 0.16.

**Frontiers**

The exact role of designability in protein evolution remains very much an open question. Although designability was first suggested in the context of convergent evolution, it is clear from the results above that this concept is by no means incompatible with divergent pictures of protein evolution. Theoretical explorations of lattice polymers has demonstrated that designability has at least the potential to leave strong marks on the course of structural evolution, and careful analysis of existing protein structures is beginning to allow us to determine what those marks might be. Our understanding of this phenomenon is, however, far from exact, and there are several outstanding questions that remain regarding this venerable idea.

One clear issue has to do with the measurement of designability on the basis of structural information. Despite the success of the ES designability definition in certain situations, this is far from a solved problem. For one, it is unclear exactly which term in Equation 2 provides the most information. In maximally compact lattice structures, for instance, the contact density (CD) is clearly exactly the same for all of the structures, a fact that does not prevent one from observing the important differences in designability between structures based on higher-order terms (32, 66, 67). In the study by Tiana *et al.* (90), the lattice proteins were not constrained to be maximally compact, and in this case the authors found that CD *decreases* over time as the polymers become less compact, but over all designability (as based on the $8^{th}$ term in the expansion) *increases* as evolution proceeds. The use of CD as a proxy for designability has yielded some interesting and important results (89, 91), but it is clear that much more exploration is needed before the convincing claim can be made that CD is a sufficient measure for the designability of a polypeptide structure. Research on this and other structural measures of designability is ongoing and can only lead to a more complete understanding of how differences in protein structure influence the size and nature of the sequence pockets that fold into those structures.

Understanding the progression of structural features such as designability over time for real proteins also requires a much more thorough understanding of the phylogenetics of protein structure. This area of study, while it is discussed at some length below, is still in its infancy, but it carries great promise. If one can build a coherent natural history of protein domains, it becomes possible to carefully measure both the changes in structure that have occurred over time (and thus determine to what extent designability has been increasing, decreasing or remaining stagnant) and also to obtain clues as to the size and extent of gene families that have been created around a given structure. Such studies should allow us to tease apart the influence of designability from the nature of divergent evolution and measure much more accurately how much of what we observe in the protein universe can be ascribed to this particular phenomenon.

**Structural Phylogeny**

**Protein Structure and Phylogenetics**

Much of the preceding discussion has focused on our attempts to understand the progress of structural evolution at varying levels of detail. As our understanding of structural evolution has progressed it has become clear that protein structural evolution, while an interesting question in its own right, may have some light to shed on the evolution of organisms themselves. As mentioned above, structural innovation certainly has important potential consequences for the space of possible functions available to an organism, and indeed one of the remaining challenges in structural evolution is understanding and quantifying this impact. Protein structure may also shed light on another important question in evolution: that of phylogenetic relationships.

An enormous amount of work has been done aimed at elucidating the phylogenies of existing and extinct organisms. In order to construct a phylogeny, one must rely on some measure of the similarity between two species coupled with the concept that organisms exhibiting a greater degree of similarity are likely to be more closely related to one another than organisms display a lesser degree of similarity. Most of these similarity measures are based on a set of observables that describe an organism (called the "characters" of that organism) and calculate evolutionary relatedness on the basis of similarity in lists of these characters. Before the advent of molecular techniques these characters were morphological in nature (92), that is, they relied on direct observations of the external or internal features of an organism. Such observations are still employed with great success today, but the introduction of molecular characters to the phylogenetic arsenal has revolutionized the study of evolutionary relationships (92-97). These molecular characters are especially important for the study of prokaryotic and microbial phylogeny since the number of available morphological characters for these organisms is small and often misleading.

The most oft-invoked molecular character was first suggested by Carl Woese and is based on the sequence of the ribosomal RNA of the small subunit of the ribosome (93, 94, 98). In this case the "structure" that is used for molecular morphology is the primary sequence of nucleotides in the molecule. Analysis of the small subunit rRNA (SSU rRNA) relationships provided some of the first evidence for the division of living organisms into three kingdoms of life (bacteria, eukarya and archaea) rather than two (prokarya and eukarya) (93-95). Indeed, this now well-accepted shift in phylogenetic categorization highlights the profound utility of molecular phylogeny: bacteria and archaea look very similar to one another if all one has at one's disposal is a set of characters one can observe in a microscope. Although molecular phylogeny has been dominated in many ways by the SSU rRNA, it is clear that phylogenetic signals may be derived from the primary structure of many genes and proteins (97). These methods rely on two (perhaps complimentary) approaches: direct sequence comparisons and gene content comparisons. The former involves computing sequence alignments and calculating distance on the basis of sequence overlap and is the method employed in constructing the rRNA phylogeny. The latter relies on definitions of orthologs (i.e. Clusters of Orthologous Groups, COGs) (99) and calculates distance based on the extent of shared orthologs. As mentioned above direct sequence comparisons are often based on a single gene like the SSU rRNA, but in some cases multiple genes are compared to one another to create a composite distance (100). Both methods exhibit certain systematic problems that have kept the question of prokaryotic phylogeny from being considered truly solved at this juncture.

Sequence comparisons exhibit a number of difficulties that revolve around the fact that the number of positions within a given sequence that can actually vary is fairly limited, and the number of sequence values that can be observed at any given position is also fairly limited (especially in the case of nucleotide sequences). Thus, on long time scales, repetitions of sequence values can be observed that obscure phylogenetic distance and thus interfere with phylogenetic signals (96). Although this is certainly a potential issue both with the rRNA trees and gene-sequence-based trees (100), perhaps more problematic is the influence of Lateral Gene Transfer (LGT, also referred to as Horizontal Gene Transfer or HGT) (75-78, 97, 101-104). LGT is a mechanism whereby genetic material is transferred from one lineage to another. This process, when it occurs between two organisms that are separated by an appreciable evolutionary distance, obscures phylogenetic signals because it leads to genetic similarity through a mechanism other than descent. The influence of LGT on both sequence-based characters and on the distribution of genes (COGs) has been considerable (75, 76), a fact that is perhaps best evidenced by the finding that many gene-specific trees differ greatly from one another and from the rRNA tree (97, 100, 101, 105-108). It has been proposed that the genetic content of some organisms, like the hyperthermophilic bacterium *Thermotoga maritima*, contain on the order of 20-25% LGT genes (78, 102). Observations such as these have lead to the proposal that a reliable phylogeny cannot be defined for prokaryotic organisms and a "web of life," rather than a tree, represents the more appropriate picture (75-77).

Although some researchers have worked to minimize or remove the influence of LGT on their gene-based phylogenies using various methods (100, 109, 110), these methods are fraught with difficulty. Selective pruning of a dataset to concentrate on a "privileged" set of genes that are less likely to be transferred (100, 111), such as genes involved in the processing of genetic information, has met with some success, but in these cases it is difficult to demonstrate *a priori* which genes are the most advantageous to use and the results are often highly sensitive to the specific set of genes that are chosen. Dutilh and coworkers recently attempted to self-consistently determine a set of these "core" genes by iteratively removing COGs from their dataset that did not agree with the overall phylogeny implied by all characters (109). Although this procedure does converge to a particular tree and gene set, over 70% of COGs are discarded during the process and it is unclear to what extent small changes in the method through which the tree is built and phylogenetic inconsistency of a given COG is determined might influence the results.

As hinted above, the distributions of protein structures in living organisms represents a potentially interesting source of phylogenetic information that is distinct from gene sequence and content and thus may not suffer from the inherent difficulties of these data sets. Perhaps the most important inherent advantage of structural information is that structural innovation is likely a very slow process and as such can provide a wealth of information about the ancient, "deep" branches of the tree of life. The emergence of a new type of protein structure quite obviously occurs on much longer time scales than microscopic sequence changes. Likewise, the discovery of a new gene can occur simply through the reorganization and recombination of existing structural domains and is thus likely to occur at a much higher frequency than true structural innovation. The above observations indicate that the transfer of a gene via LGT is not guaranteed to involve the transfer of a new structural domain from the donor lineage to the acceptor lineage. The slower pace of structural innovation guarantees that two organisms are likely to be more structurally similar than genetically similar across any given evolutionary distance, and so the probability of transferring a new domain *via* LGT is clearly much lower than the probability of transferring a new gene. Although LGT could certainly result in a structural content change for the acceptor organism, *a priori* it is unclear to what extent this process has interfered with the structural signal. These potential advantages that derive from considering the protein structural characters of an organism have underpinned a number of attempts to use structural information to construct phylogenies.

**Building Structural Phylogenies**

Early work on structural phylogeny revolved around comparing the fold content of various organisms. The first attempt at this was made by Koonin and coworkers as part of an effort to classify structural domains into the three kingdoms of life (79). The phylogeny they produced as part of this work involved only a very few taxa, and the phylogeny suggested by their data at that time was not terribly reasonable. Later, Lin and Gerstein extended this work by employing an increased number of available structures to an increased number of taxa. Despite this improvement, however, the phylogenies produced by a variety of methods from the fold-content dataset were also somewhat unreasonable and did not agree in many basic ways with much of the other sources of phylogenetic signals available at the time (112). A similar study in 2003 by Caetano-Anollés and Caetano-Anollés recovered the Eukaryote/Bacteria/Archaea organization but did not discriminate well the overall organization of taxa within the domains themselves (113). The mediocre performance of structural characters in these studies most likely results from the fact that the "fold" level of structural classification is too coarse-grained to provide many benefits over gene sequence-or content-based datasets. For instance, two organisms may actually share very few domains but, since they have at least one (separate) representative from the same set of folds, exhibit perfect overlap at the fold level. The fact that there are fairly few folds that had (or have) been described (6), combined with the observation that many of these folds are very widely conserved (6, 79, 80), indicates that fold innovation may be *too* slow and thus contain few useable phylogenetic signals. It is important to note that this earlier work on structural phylogeny involved very few taxa, so it is possible that the difficulties encountered in those studies were the result not only of the very coarse-grained nature of the fold description but also the limited taxon sampling available at the time (79, 112, 113).

More recent and (arguably) more successful phylogenies have been built on the basis of protein structural domains as the fundamental set of characters. One of the clearest advantages of structural domains is that they represent specific sequence-structure pairs and are very unlikely to have ever evolved convergently. Current theoretical models of domain evolution also provide an important and unprecedented advantage to this particular data set. All methods for phylogenetic reconstruction have at their core a set of assumptions and algorithms that allow one to translate similarity in some set of characters into evolutionary distances. Oftentimes one chooses these algorithms on the basis of how "reasonable" they seem for the set of characters in question and how computationally intensive they are. In the case of structural domains, however, it is possible to start from an independently testable model of the evolution of those characters and attempt to translate that model into a phylogenetic method (or set of methods) that most closely correspond to the nature and assumptions of that model.

The clearest example of this difference has to do with understanding the prevalence of LGT. In the speciation model of structural evolution discussed in the structural proteome section above, transfer of domains between one lineage and another never occurs (74). This model reproduces the observation that structural proteomes are highly non-random subsets of the PDUG. When one looks at the structure of this model, it becomes apparent that the partitioning of the graph into specific proteomes that do not intermingle is basically the cause of this behavior—in a sense, each lineage-specific PDUG is a separate, small universe of its own. Lateral Structural Domain Transfer (LSDT) resulting from LGT in a sense mimics convergent node discovery, and as such this result points to the working hypothesis that LSDT might not have been widespread in all lineages. Indeed, when a variety of LSDT mechanisms are implemented into the evolutionary model, they tend to increase the probability that model organismal subgraphs are random subsets of their model structural universe (74, 114). Although these results certainly in no way constitute proof that LSDT has been rare, they point to a particular set of phylogenetic methods that are more consistent with this finding. Given that no analogue of the PDUG exists for sequence- or gene-based characters, this type of analysis is currently limited to structural information.

The method that is perhaps most consistent with these structural evolution models is known as Dollo parsimony (115, 116). Methods that are based on Maximum Parsimony (MP) build trees by modeling the gain, loss, and transformation of characters at different internal nodes on the tree. For a certain tree topology, the most reasonable evolutionary scenario is considered to be the one in which the fewest events are necessary to explain the existing distribution of characters (117). The best topology is then also taken to be the one that requires the fewest events in its "best-case" scenario. There are a variety of MP methods available (97, 112, 116, 117), and as mentioned above the most reasonable method for structural domains (based on the assumption that convergent discovery of domains and LSDT are both very infrequent) is Dollo. This type of parsimony involves a constraint (called the Dollo constraint) that prevents "reversions" on a tree. This means that any time a lineage looses a particular character (i.e. a particular domain), it cannot be rediscovered at a later point in that lineage. This stipulation effectively ensures that every single character on the tree is monophyletic, that is, has only one point of origin in the phylogeny (115, 116). This is in contrast to "unconstrained" MP methods that allow for any degree of homoplasy—that is, methods that allow for the loss and gain of a particular character at will. Analysis of these two methods clearly indicates that Dollo is the more reasonable assumption in the light of current structural evolution models; in fact, in the limit where those models represent an exact picture of structural evolution the Dollo tree will represent an exact solution of the phylogenetic problem (74, 114).

Starting with the structural proteomes consisting of PDUG domains as described above (33, 74, 81, 114), we built a prokaryotic phylogeny on the basis of both MP methods. The Dollo constraint provided a remarkably reasonable phylogeny that, on balance, supported many of the major groupings indicated by the rRNA phylogeny and various other molecular methods (97, 114). In some cases where the Dollo phylogeny differs from "canonical" groupings well supported by other methods, the source of difficulty was clear: a greater probability that the oPDUG is a random subgraph of the PDUG. For instance, in the Dollo tree both *Campylobacter jejuni* and *Helicobacter pylori* are moved from their traditional placement in the proteobacterial clade to a position near the base of the tree along with the cyanobacteria (97, 114). These proteomes, however, have among the highest probabilities of being random subgraphs within the dataset we employed, and it is clear that this degree of randomness interferes with the phylogenetic signal. This observation highlights the utility of our domain-based approach as we were able to identify a source of difficulty in the placement of these organisms in this and perhaps other phylogenies (114). Other non-traditional groupings that are observed in the Dollo tree, however, occurred with proteomes that have a low probability of being random. The grouping of *T. maritima* with *Thermoanaerobacter tengcongensis* in the gram-positive clade is one example. In this case the lack of randomness indicates that the structural signal may be contributing signals that are difficult to obtain from other datasets and represent potentially important phylogenetic hypotheses that warrant further study.

When we employed unconstrained parsimony, however, the results were not as reasonable as those obtained using the Dollo method. Although there was some correspondence between the trees produced by both methods, the unconstrained method splits the gram-positive clade of bacteria and asserts that the high G+C subset of this clade is actually more closely related to the proteobacteria than to the remainder of the gram positives. Such a grouping is very rare in other phylogenetic methodologies and is simply less phylogenetically reasonable (97, 100, 101, 105-107, 114), providing a measure of phylogenetic support for the claim that LSDT has played a minor role in the evolution of structural proteomes and the Dollo constraint is necessary to elicit a good phylogenetic signal from structural characters using MP.

Another method that is consistent with current theoretical models is the Neighbor Joining (NJ) method (116, 118). This method uses a calculated distance between each pair of taxa to build the tree; in the structural case this distance is just the number of characters in which the organisms differ (also referred to as the "Hamming Distance" in the sequence of descriptors (112)). Although the NJ algorithm does not explicitly prevent convergence or LSDT as in the case of Dollo parsimony, in the limit where current models of structural evolution are exact the distance-based NJ method should give results roughly equivalent to those obtained from Dollo parsimony. This method does indeed produce a phylogeny that is very similar to that based on Dollo parsimony, although in this case the statistical support for some of the "important" internal nodes is less strong than in the Dollo tree (114).

Many of the conclusions of our work summarized above were recently confirmed by the work of Bourne and coworkers (119) based on domains classified in SCOP rather than on FSSP domains represented by the PDUG. These authors employed a single phylogenetic method (NJ) and could not make statements about the randomness of proteomes given that the SCOP database lacks an analog of the PDUG from which P-values can be calculated. Bourne and coworkers benefited, however, from a larger set of taxa, which allowed them to expand their analysis to include eukaryotes in addition to bacteria and archaea. Close comparison of their bacterial tree and ours reveals almost perfect agreement, which, given the similarity in datasets, is not surprising (114, 119). For instance, their tree places *H. pylori* and *C. jejuni* with the cyanobacteria outside of the proteobacterial clade and *T. maritima* with *T. tengcongensis* as is seen in our Dollo and NJ trees.

One interesting question that arises in analysis of the above work involves understanding exactly why protein structural domains are less likely to be influenced by lateral gene transfer (as evidenced by phylogenetic "success" of two completely independent methods based on unpruned datasets). At least in the bacterial clade, part of the success may simply derive from the fact that the domain overlap of even phylogenetically distant organisms is considerable—in the case of *Escherichia coli* and *Bacillus subtilis*, for instance, 873 of the 959 domains from *B. subtilis* are shared between the two organisms (74). This finding implies that an LGT event over a considerable evolutionary distance has a fairly low probability (~10% if domains are sampled equally) of resulting in a change in the structural content of either organism.  Consistent with this picture, Bourne and coworkers found that consideration of the abundance of domains, rather than just the presence or absence of structural information, provided phylogenies of much lower quality (119).  Although the authors attribute this fact to gene duplication mechanisms, it is far more likely to be a reflection of the influence of LGT.  LGT events that do not result in the transfer of new structural domains to the acceptor lineage will, by definition, only influence the abundance of existing domains and thus the widespread influence of LGT observed in gene-based phylogenies will be recapitulated in that kind of structural dataset.

The results presented above also provided further evidence that a great deal of structural information is widely conserved even among the three kingdoms of life (79, 112, 119).  Given the large number of structural domains that are very ancient, it is interesting to speculate on the tenor of structural evolution that predated the true LUCA of all cellular life.  Although the inherent lack of phylogenetic information before this point makes study of this process inherently difficult, analysis of the results discussed above has some surprising implications.  For instance, the number and diversity of domains that likely existed in the LUCA organism indicates that this organism is likely to have been surprisingly structurally sophisticated.  All major classes of protein structure (all-$\alpha$, all-$\beta$, $\alpha+\beta$ and $\alpha/\beta$) are represented in this set.  Given the short period of time in which life evolved before the split of the three kingdoms (perhaps 500 million years or so), it is clear that the rate of structural innovation in this time period was quite considerable.  Understanding the nature of structural evolution in these early stages, and the mechanisms or conditions that might have separated it from the ensuing structural evolution observed within each of the three kingdoms, is an intriguing question that requires a great deal of further study.

**Frontiers**

The promise of structural phylogeny is, as yet, largely unrealized for a variety of reasons. Foremost among these is the fact that experimental exploration of the evolved structural universe is paltry compared to sequence exploration (6, 21, 24, 120). Although it has been claimed that that about 50% of "existing" sequences exhibit sequence homology with structures that have been characterized (80), this is a very difficult statement to interpret. Despite the growing number of fully-sequenced genomes, it is clear that our taxon sampling even on the sequence level is nowhere near complete, especially within the bacterial and archaeal kingdoms where there are many species that have not been cultured and are known only by a single 16S rRNA sequence. If and when some appreciable fraction of this organismal diversity has been sampled in the set of fully sequenced genomes, it will be possible to more accurately asses to what extent our exploration of the structural universe is complete at the domain level. Regardless of this fact, the structural information employed in the phylogenetic work described above is by no means a complete set. This lack of completeness may also lead to biases in this character set that may impact the phylogeny, especially within the archaeal kingdom where structural coverage is fairly low and exact relationships are thus difficult to determine (114, 119). Structural genomics projects may alleviate some of these difficulties by providing at least a few glimpses at structural proteomes that are as unbiased as possible, but a great deal of structure determination will be necessary to provide a higher-quality dataset from which to build structural phylogenies. As the characters currently stand, they could still provide a great deal of further information towards the phylogenetic effort, especially if combined with other types of data in a hybrid approach.

Given that domain-based structural characters have proven somewhat reliable in phylogenetic reconstructions, it is now possible to consider building a natural history of structural domains on the basis of the species tree (114, 117, 119). If successful and reasonable, such efforts could provide an immeasurable amount of information about the evolution of protein structures and allow for the detailed testing of evolutionary models. The main difficulty with this particular line of inquiry is that the results are highly dependent on the topology of the tree itself (for which no consensus exists) and on the assumptions that describe the evolution of the characters (117). Both of these problems can be overcome by considering a "self-consistent" approach where structural information is used to build the (evolutionarily reasonable) tree itself. This effort should move our understanding of structural evolution beyond the simple considerations of widespread conservation (79, 80, 112, 119). As our picture of the natural history of domains themselves improves, more accurate models of structural evolution may be posed and tested against the features of that natural history. In essence this may produce a feedback cycle in which ever-more reasonable pictures of structural evolution inform ever-more accurate phylogenetic scenarios.

## Conclusions

Protein structural evolution is a fascinating subject and has been the object of intense study ever since it became clear that proteins had specific structures that could be specified on the basis of sequence alone. The intense pace of structural biology and the corresponding explosion in structural information has supported much of the work outlined above, and as mentioned earlier it is clear that further structural exploration, especially from structural genomics projects, will only increase the precision with which quantitative statements about structural evolution may be made.  Despite the volume of work that has been poured into the question of protein structural evolution and the number of researchers who have contributed to developing the structure-centric view outlined in this chapter, this work has allowed the community to pose an increasingly detailed set of questions.  Answering these questions (some of which were posed in the "Frontiers" sections above) will require intense theoretical work, deep evolutionary thinking, and most of all sustained experimental effort to provide the greatest amount of structural information possible.

## Acknowledgements

The authors thank Dr. Eugene Koonin, Dr. Nikolay Dokholyan, and C. Brian Roland for many illuminating and helpful discussions.